%% file: bare_adv.tex
\documentclass[10pt,journal,compsoc, twoside]{IEEEtran}

\ifCLASSOPTIONcompsoc
  \usepackage[nocompress]{cite}
\else
  \usepackage{cite}
\fi
\ifCLASSINFOpdf
\else
\fi
\usepackage{amsmath}
\usepackage{ragged2e}
\usepackage{amsfonts,amssymb}

\usepackage{makecell}
\usepackage{footnote}
\allowdisplaybreaks[4]
\usepackage{mathrsfs}
\usepackage{caption}
\usepackage{amsthm}
\usepackage{amsmath} 
\usepackage{graphicx}
\usepackage{booktabs} 
\usepackage[linesnumbered,ruled,vlined]{algorithm2e}
\usepackage{url}
\usepackage{ctable}
\usepackage{multirow}
\usepackage{algpseudocode}
\usepackage{pifont}
\usepackage{color}
\usepackage{subfigure}
\usepackage{makecell}
\usepackage{bm}
\usepackage{enumitem}
\usepackage{caption}
\usepackage{cite}
\setlist{leftmargin=10pt}
\begin{document}
%
\title{A Survey of Community Detection Approaches: From Statistical Modeling to Deep Learning}

\author{Di Jin, Zhizhi Yu, Pengfei Jiao, Shirui Pan, Dongxiao He, Jia Wu,\\ Philip~S.~Yu
and Weixiong Zhang
\IEEEcompsocitemizethanks{\IEEEcompsocthanksitem D. Jin, Z. Yu, P. Jiao and D. He are with Tianjin University, China. 


\IEEEcompsocthanksitem S. Pan is with Monash University, Australia. 

\IEEEcompsocthanksitem J. Wu is with Macquarie University, Australia. 

\IEEEcompsocthanksitem P. S. Yu is with University of Illinois at Chicago, USA.

\IEEEcompsocthanksitem W. Zhang is with Washington University, USA.


\IEEEcompsocthanksitem Corresponding author: Pengfei Jiao.
}
}

%

\IEEEpeerreviewmaketitle

\IEEEpubid{xxxx-xxxx/0x/\$xx.00 © 200x IEEE \qquad Published by the IEEE Computer Society}

\IEEEtitleabstractindextext{%
\begin{abstract}
\justifying
Community detection, a fundamental task for network analysis, aims to partition a network into multiple sub-structures to help reveal their latent functions. Community detection has been extensively studied in and broadly applied to many real-world network problems.
Classical approaches to community detection typically utilize probabilistic graphical models and adopt a variety of prior knowledge to infer community structures. As the problems that network methods try to solve and the network data to be analyzed become increasingly more sophisticated, new approaches have also been proposed and developed, particularly those that utilize deep learning and convert networked data into low dimensional representation.
Despite all the recent advancement, there is still a lack of insightful understanding of the theoretical and methodological underpinning of community detection, which will be critically important for future development of the area of network analysis.
In this paper, we develop and present a unified architecture of network community-finding methods to characterize the state-of-the-art of the field of community detection. Specifically, we provide a comprehensive review of the existing community detection methods and introduce a new taxonomy that divides the existing methods into two categories, namely probabilistic graphical model and deep learning. We then discuss in detail the main idea behind each method in the two categories. Furthermore, to promote future development of community detection, we release several benchmark datasets from several problem domains
and highlight their applications to various network analysis tasks. We conclude with discussions of the challenges of the field and suggestions of possible directions for future research.
\justifying
\end{abstract}

}

\maketitle

\IEEEdisplaynontitleabstractindextext

%
\IEEEpeerreviewmaketitle

\ifCLASSOPTIONcompsoc


\maketitle
\input{introduction}

\input{problem_definition}

\input{statistical_modeling}

\input{deep_learning}

\input{Application}

\input{challenges}
\input{future}

\section{Conclusion}
In this paper, we provide a comprehensive and up-to-date
literature review on the community detection approaches. 
One of our main objectives is to organize and present most of the work conducted so far in a unified perspective. In a first step, we discuss in details the problem of community detection, and provide a new taxonomy to group most existing methods into two categories from the perspective of learning: probabilistic graphical model and deep learning. We then thoroughly review, compare, and summarize the existing methods in these two categories, and discuss how some of these methods can be interested. 
Moreover, since the problem is highly application oriented, we introduce a wide range of applications of community detection in various fields. We also highlight that more effort is needed to address several challenging open problems for the research of community detection. 
Last but not least, while we cannot list all the literature on community detection, we expect that our view that attempts to synthesize the state-of-the-art 
of community detection will contribute to a better understanding of this highly active and increasingly important area of study in network science, serve as a source of information for new researchers entering this field and the researchers 
working in this area, and promote future developments of next-generation community detection approaches.




\ifCLASSOPTIONcaptionsoff
  \newpage
\fi


\bibliographystyle{ieeetr}
\bibliography{reference.bib}
\clearpage
\appendices
\section{}
Here we list the key terms and notations in the main text in Table 1.
\begin{table}[h!]
	\centering
	 \caption{\label{tab:notations} Summary of notations.}
	\scalebox{0.99}{
	\begin{tabular}{p{1.25cm} p{6.5cm}}
		\toprule
		{\bf Notations}      & {\bf Descriptions}\\
		\midrule
		$G$ & A network. \\
		$V, E$ & The sets of nodes and edges of a network. \\
		$A, X$ & The adjacency matrix and node attribute matrix.\\
	    $D$ & The node degree matrix. \\
        $n, m$ & The numbers of nodes and edges.\\
        $e_{ij}$ & The edge between nodes $v_i$ and $v_j$.\\
        $a_{ij}$ & The connection between nodes $v_i$ and $v_j$.\\
        $x_{i}$& The attribute vector and degree of node $v_i$.\\
        $q$& The maximal number of node attributes.\\
		\hline
		$\mathcal{C}$ & The set of communities.\\
		$C$ & The community assignments of nodes.\\
		$k$ & The numbers of communities.\\ 
		$c_{i}$& The community which node $v_i$ belongs to.\\
		$\omega_r$ &The probability of nodes assigned to community $\mathcal{C}_r$.\\
		$\pi_{rs}$ &The probability of link generation within two communities $\mathcal{C}_r$ and $\mathcal{C}_s$.\\
		$\delta (c_i, c_j)$& The probability of nodes $v_i$ and $v_j$ falling into the same community partition. \\
		\hline
		$E(C;A)$& The energy function in MRF.\\
        $\Theta _{i}$& The unary potential function in $E(C;A)$.\\
        $\Theta _{ij}$& The pairwise potential function in $E(C;A)$.\\
		\hline
 		$KL(\cdot||\cdot)$ & KL-divergence. \\
		$\widetilde{A}$ & The estimation matrix of $A$ in NMF.\\
		$B, S$ & Community membership
		matrix and attribute community matrix in NMF. \\ 
		$H, W$ & Node representation matrix and weight matrix of neural networks. \\
		$M, L$ & Modularity matrix and Laplacian matrix. \\
 		$\mathcal{F}$ & The set of factor nodes. \\ 
		$\widehat{A}, \widehat{X}$ & Reconstructed adjacency matrix and node attribute matrix.\\
		$\mathcal{G}, \mathcal{D}$ & The generator and discriminator of GAN. \\ 
		$\mathcal{E}$ & The encoder that derives node representation. \\ 
		\bottomrule
	\end{tabular}
	}
	\vspace{0cm}
\end{table}

\section{}
In Section 3.1.1, we have introduced several  SBM variants for community detection.
Here, we give the overall process of community detection based on the basic SBM with a Bernoulli distribution \cite{Holland1983Stoch}, MMSB \cite{Airoldi2008Mixed} and DSBM \cite{DBLP:journals/ml/YangCZGJ11} in algorithm 1, algorithm 2 and algorithm 3 respectively.

\begin{algorithm}[h!]
\label{1basicsbm}
\caption{The basic SBM-based method \cite{Holland1983Stoch}}
\KwIn{$n$, $k$.} 
\KwOut{the community assignments of nodes $C$.}
\vspace{2mm}
Assume that nodes are independently divided into $k$ communities\;
Inference the parameters $\omega$, $\pi$ of likelihood function by using EM algorithm\;
\For{each node $v_i$}{
\For{each node $v_j$}{
$c_i \overset{i \cdot i \cdot d}{\sim} \operatorname{Multinomial} (1; \omega)$\;
$a_{ij}|c_{ir}, c_{js} \overset{i \cdot i \cdot d}{\sim} \operatorname {Bernoulli} (\pi_{rs}) | 0<r,s\leq k$\;
}
}
\Return $C$\;
\end{algorithm}

\begin{algorithm}[h!]
\label{MMSB}
\caption{The MMSB-based method \cite{Airoldi2008Mixed}}
\KwIn{$n$, $k$, $\alpha$, $\beta$.}
\KwOut{the community assignments of nodes $C$.}
\vspace{2mm}
\For{each node $v_i$}{
$\omega_i \sim \operatorname{Dirichlet}(\alpha)$\;
Assigns the community assignments $c_i$ with $\omega_i$\;
\For{each node $v_j$}
{

$\pi_{rs}|c_{ir}, c_{js} \sim \operatorname{Beta}(\beta) | 0<r,s\leq k$\;
$a_{i \rightarrow j} \sim \operatorname{Multinomial} (\omega_i)$\;
$a_{i \leftarrow j} \sim \operatorname{Multinomial} (\omega_j)$\;
$a_{ij} \sim \operatorname{Bernoulli}( a_{i \rightarrow j}^T \pi_{rs} a_{i \leftarrow j})$\;}
}
\Return $C$\;
\end{algorithm}

\begin{algorithm}[h!]
\label{dSBM}
\caption{The DSBM-based method \cite{DBLP:journals/ml/YangCZGJ11}}
\KwIn{$n,\pi,A,T$.} 
\KwOut{the community assignments of nodes $C^{(T)}$.}
\vspace{2mm}
\If{time $t == 1$}{generate the social network followed by SBM\;}
\For{each time $t>1$}{generate $c_i^{(t)} \sim \pi (c_i^{(t)}| c_i^{(t-1)},A)|0<i \leq n$\;}
\For{each pair of nodes $(v_i,v_j)$ at time $t$}{generate $w_{ij}^{(t)} \sim \operatorname{Bernoulli}(\cdot|\pi_{c_i^{(t)}, c_i^{(t-1)}})|0< i,j \leq n$\;
}
\Return $C^{(T)}$\;
\end{algorithm}

\section{}
In Section 3.1.2, we have  presented several  topic models for community detection. Here, we give the generation process of SSN-LDA \cite{DBLP:conf/isi/ZhangQGFY07} for one social interaction profile in algorithm 4, and provide the process for clustering attribute communities \cite{xu2012model} in algorithm 5.
\begin{algorithm}[h!]
\label{SSNLDA}
\caption{The SSN-LDA method \cite{DBLP:conf/isi/ZhangQGFY07}}
\KwIn{$k, \vec{\alpha}, \vec{\beta}, \varepsilon.$}
\KwOut{the community assignment of one node $c_i$.}
\vspace{2mm}
sample mixture components $ \vec{\phi} \sim \operatorname{Dirichlet}(\vec{\beta})
$\;
choose $\vec{\theta_i} \sim \operatorname{Dirichlet}(\vec{\alpha})$\;
choose $N_i \sim \operatorname{Poisson}(\varepsilon)$\;
\For{each neighbor $v_j$ of $v_i$} {
choose a community $c_{i} \sim \operatorname{Multinomial}(\vec{\theta})$\;
choose a social interaction $a_{ij} \sim \operatorname{Multinomial}(\vec{\phi_{c_{i}}})$;}
\Return $c_{i}$\;
\end{algorithm}
\clearpage
\begin{algorithm}[h!]
\label{1da+sbm}
\caption{The generation of Bayesian attributed graph clustering (BAGC) \cite{xu2012model}}
\KwIn{$n, k, T, \mathcal{C}$, a time attributes set $\Lambda=\{\lambda^{(1)}, \ldots ,\lambda^{(T)}\}$, parameters $\varepsilon, \mu, \nu$.}
\KwOut{the community assignments of nodes $C$, attribute matrix $X$.}
\vspace{2mm}
choose $\alpha \sim \operatorname{Dirichlet}(\varepsilon)$\;
\For{$\mathcal{C}_i \in \{\mathcal{C}_1,\mathcal{C}_2,...,\mathcal{C}_k\}$} 
{
	\For{each attribute $\lambda^{(t)}$}{
			choose $\theta_i^{(t)} \sim \operatorname{Dirichlet}(\lambda^{(t)})$\;}
	\For{each community $\mathcal{C}_j \in \{\mathcal{C}_i,\mathcal{C}_{i+1},...,\mathcal{C}_k\}$}{choose $\phi_{ij} \sim \operatorname{Beta} (\mu, \nu)$;
			}

}
\For{each node $v_{i}$}
{
	choose $c_i \sim \operatorname{Multinomial}(\alpha)$\;
	\For{each attribute $\lambda^{(t)}$}{
			choose $x_i^{(t)} \sim \operatorname{Multinomial}(\theta_{c_i}^{(t)})$\;}
	\For {each node $v_j$ with $i>j$}{
			choose $a_{ij}\sim \operatorname{Bernoulli}(\phi_{c_i c_j})|0<i,j \leq n$\;}
}
\Return C, X\;
\end{algorithm}
\end{document}

%% file: introduction.tex
\section{introduction}
\begin{figure}[htbp]
\centering
\setlength{\abovecaptionskip}{0.05cm}
\setlength{\belowcaptionskip}{-0.35cm}
\includegraphics[width=0.99\linewidth]{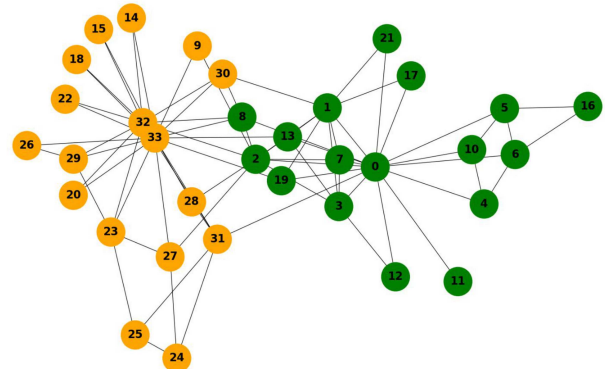}
\caption{An illustrative example (Zacharys karate club network \cite{Wayne1977An}) showing community structure. The nodes of this network are divided into two groups, with most connections falling within groups and only a few between groups.}
\label{karate}
\vspace{-1mm}
\end{figure}

\IEEEPARstart{N}{etwork} science is the study of complex systems in the form of networks using theories and techniques of computer science, mathematics and physics. In particular, network structures \cite{2002Girvan} (see an example in Fig \ref{karate}) have been studied extensively under the notions of subgraphs, network modules, and communities. Identification of network structures or community detection is to divide nodes in a network into groups where the nodes in a group are densely connected whereas nodes in different groups are sparsely linked. Mining network structures is also the key to revealing and comprehending organizational principles and operational functions of complex network systems. For example, community detection has been applied to 
recommendation \cite{DBLP:conf/kdd/SatuluriWZQWDTJ20, DBLP:conf/icdm/MukherjeeLW15}, anomaly detection \cite{DBLP:journals/ijeb/KeyvanpourSG20, DBLP:journals/tcns/WangP17}, and terrorist organization identification \cite{DBLP:conf/rcis/SaidiTG18},
just to name a few. 
Much effort has also been devoted to the analysis of network structural
properties, e.g., the small world effect (i.e., the average distance between nodes is short \cite{1998Collectivedynamics}) and the scale free property (i.e., the distribution of node degrees follows a power law distribution \cite{Barab1999Emergence}).  


Many algorithms for community detection have been proposed, a majority of which employ exclusively the information of network topology. They include hierarchical clustering \cite{DBLP:series/ads/2010-40} \cite{2015Defining}, modularity optimization \cite{DBLP:conf/ijcai/YangCHWWZ16} \cite{2004Finding} \cite{DBLP:journals/pnas/ZhangM14}, spectral clustering \cite{2017Magnetic} \cite{DBLP:conf/www/LiHBH15} and statistical inference \cite{DBLP:journals/jmlr/AnandkumarGHK14} \cite{DBLP:conf/aaai/HeLJZ15}. 
New methods were developed to utilize node semantics or node attributes in addition to network topology to improve the quality of resulting communities and meanwhile provide explanation to the results. 
These include heuristic optimization (multi-objective) \cite{2019Multiobjective}\cite{DBLP:journals/tcyb/LiLW18}, matrix factorization \cite{wang2011community} \cite{zhang2012overlapping} and Bayesian model \cite{DBLP:conf/aaai/0002ZL15}.
As more complex network problems were tackled, complex network data from multiple sources, e.g., network topology and node semantics, must be effectively integrated. As a result, it became difficult for these traditional approaches to perform data fusion effectively on data of very high dimensions and diverse properties.
The technique of deep learning was recently adopted to handle the high dimensional network data and learn low-dimensional representation of network structures. Examples include methods based on the auto-encoder \cite{wang2017mgae:} \cite{sun2017a} and the generative adversarial approach \cite{1} \cite{DBLP:conf/kdd/ZhangXYLWZY20}. 

An important and effective idea for community detection is to learn an adequate representation of the network structure of a given network.
We call such approaches
{\em learning-based community detection}. 
Among these methods are the model-based generative models. The most popular and representative example is the stochastic block model (SBM) \cite{Holland1983Stoch}, which detects communities by formalizing a generative process of a network as a sequence of rigorous probability distributions. 
Several extensions and improvements have been introduced to boost the performance of SBM \cite{Airoldi2008Mixed}\cite{dcsbm}. 
Another model-based learning approach adopts Markov random field (MRF), an undirected graphical model, to take advantage of neighborhood structures in networks \cite{He2018}.
A primary recent development in learning-based methods exploits the low-dimensional representation capability of deep learning. For instance, convolutional neural network (CNN) \cite{19} utilizes
convolution and pooling operations to reduce the dimensions of network data, 
so as to effectively discover communities in 
networks. Graph convolutional network (GCN) \cite{DBLP:conf/iclr/KipfW17}, which inherits the advantages of CNN and directly operates on network structured data, has also been explored to derive community representation \cite{WWW}. 

Despite the endured effort to develop effective methods for community detection \cite{DBLP:conf/icde/YangCZWPC18, 
DBLP:journals/tkde/JinWHDZ21}, there is still a lack of understanding of the theoretical and methodological underpinning of community detection, particularly that based on learning. To
bridge this gap, in this paper, we will provide a synthesized survey of the existing representative methods. We focus particularly on two general lines of approaches, one based on probabilistic modeling 
and the other on deep learning. We start with a detailed description of each line of the work and provide a thorough review and comparison of the methods. We then consider several applications of community detection in diverse fields. We finish with the discussion of some critical challenges of the field of network analysis and directions for future rewarding research.

One of the major objectives of our survey is to provide a new perspective on the existing methods to help better understand the fundamental issues and enabling techniques for community detection. Our survey differs from the published ones in three aspects.
First, we summarize the existing methods by focusing on learning, a central issue of community modeling and community detection, whereas the existing reviews \cite{2013Overlapping} \cite{DBLP:journals/csur/AggarwalS14} generally discuss the chronological development of the existing methods. 
Second, we present a recent trend in the development of methods for community detection, i.e., from statistical modeling to deep learning,
while the others focus mainly on individual techniques, e.g., evolutionary computation \cite{DBLP:journals/tec/Pizzuti18}, statistical inference 
\cite{DBLP:journals/jmlr/Abbe17} or deep learning \cite{DBLP:conf/ijcai/LiuX0ZHPNYY20}. 
Third, we present a unified system architecture to characterize the existing methods, which provides 
a novel and 
synthesized perspective on 
methods based on statistical modeling and deep learning, which go significantly beyond some of the existing surveys \cite{2013Clustering} \cite{DBLP:series/lncs/HartmannKW16}. Last but not least, in the era of deep learning, as network data become increasingly more complex and various ideas and techniques have been proposed, a survey is urgently needed to comprehensively unravel the inherent relationships among the existing methods for community detection. 

Aiming at offering a general guidance to researchers and practitioners who are interested in network science and network data analysis, we make our unique contributions in this work as summarized below.
\begin{itemize}
    \item We present the most comprehensive and extensive overview of learning-based community detection and divide them into two categories,  
    probabilistic graphical model and deep learning. To the best of our knowledge, this is the first attempt devoted to community detection from the perspective of learning. It offers a solid foundation for understanding the intuition behind community detection, and can be used as a guideline for designing and utilizing different methods for community detection.
    \item We provide a thorough theoretical analysis of learning-based community detection methods, discuss their similarities and differences, identify critical challenges that remain poorly addressed and point out five directions for future development.
    \item We gather abundant resources on learning-based community detection, including state-of-the-arts benchmark datasets and applications. 
\end{itemize}

The rest of this survey paper is organized as follows. Section 2 gives the preliminaries and categorization of existing community detection approaches. Section 3 presents a technical overview of research progress in statistical modeling for community detection. Section 4 overviews the research on deep learning-based community detection. Section 5 discusses applications of community detection. We suggest promising future research directions in Section 6 and conclude in Section 7.

%% file: problem_definition.tex
\section{Preliminaries and Categorization}
We first introduce the terms and notations, and then present a classification of the methods for community detection that we will discuss in this paper.

\subsection{Definitions, Terms and Notations}


\begin{figure*}[htbp]
\centering
\setlength{\abovecaptionskip}{0.3cm}
\setlength{\belowcaptionskip}{-0.3cm}
\includegraphics[width=0.98\linewidth]{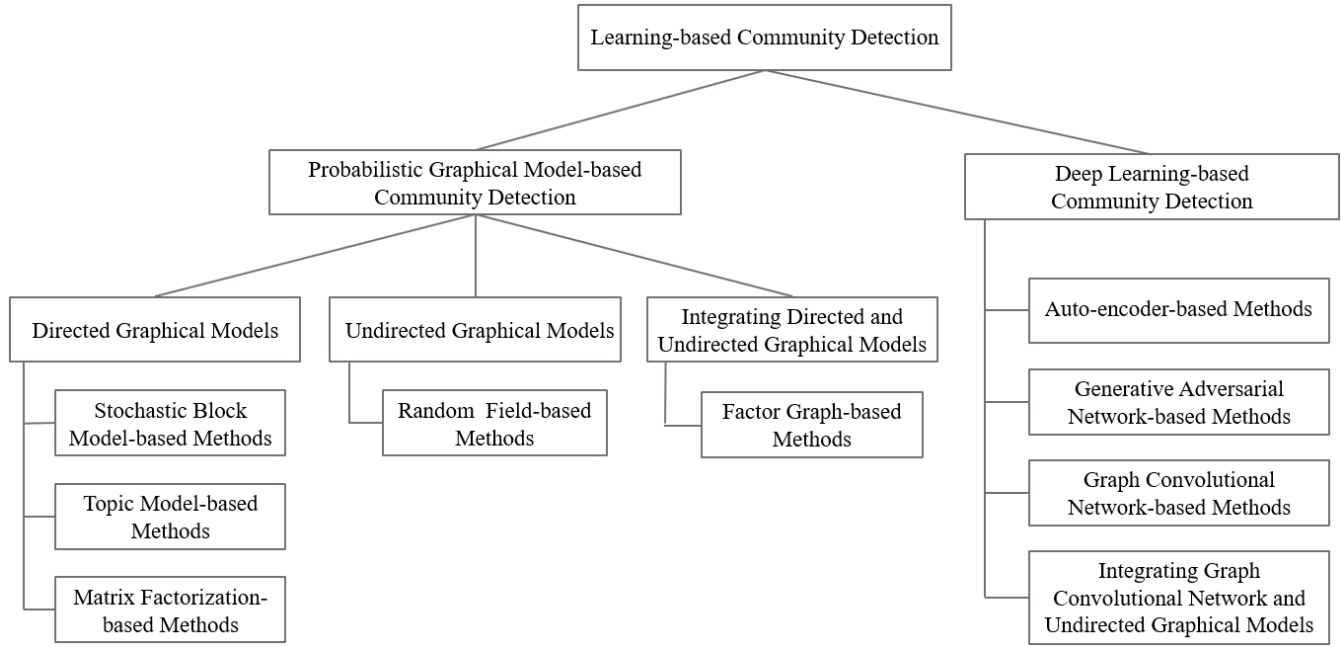}
\caption{Classification breakdown of methods for community detection.}
\label{overview1}
\end{figure*}

\emph{Definition 1.} \textbf{Network}. A network $G = (V, E, X)$ consists of $n$ nodes $V=\left\{v_{1}, v_{2}, \ldots, v_{n}\right\}$, $m$ edges $E= \{e_{ij}\} \subseteq V \times V$, and a maximal number $q$ of attributes $x_i$ on a node $v_i$, where all $x_i$'s collectively give rise to an $n\times q$ attribute matrix $X = (x_{i})_{n\times q}$. The topological structure of $G$ can be defined by an $n\times n$ adjacency matrix $A = (a_{ij})_{n\times n}$, where $a_{ij} = 1$ if ${e_{ij}} \in E$, or 0, otherwise. $G$ is undirected if $a_{ij} = a_{ji}$, or directed, otherwise \cite{2013Clustering}. 


\emph{Definition 2.} \textbf{Community}. The network $G$ contains $k$ communities $\mathcal{C} = {\left\{\mathcal{C}_{1}, \mathcal{C}_{2}, \ldots, \mathcal{C}_{k}\right\}}$, where $\mathcal{C}_i$ is a subgraph of $G$ and the nodes within $\mathcal{C}_i$ are densely connected whereas the nodes across $\mathcal{C}_i$ and $\mathcal{C}_j$ are sparsely connected.
The communities are non-overlapping when $\mathcal{C}_{i} \cap \mathcal{C}_{j} = \emptyset \,\, {\forall i, j}$.

\emph{Definition 3.} \textbf{Community Detection}. Given a network $G$, community detection is to design a mapping $\mathcal{F}$
to assign every node $v_i$ of $G$ into at least one of the $k$ communities, i.e., to label $v_i$ at least one community identity $c_i \in {\left\{\mathcal{C}_{1}, \mathcal{C}_{2}, \ldots, \mathcal{C}_{k}\right\}}$. Equivalently, the problem is to derive a community assignment of nodes $C=(c_{1}, c_{2}, \ldots, c_{n})$.

\subsection{A Taxonomy of Community Detection Approaches}
To help better understand the existing learning-based methods and facilitate our discussion in the rest of the paper, we introduce a taxonomy of the methods for community detection (Fig. \ref{overview1}) where the methods are grouped into two categories, probabilistic graphical model and deep learning.

    \textbf{Probabilistic graphical model-based methods} employ heuristics or meta-heuristics on community generation to discover network communities. These methods typically adopt some models of network structures to describe the dependencies among the entities (i.e., nodes) via the edges of the networks. Depending the type of probabilistic graphical models used, community detection can be further divided into three main categories: directed graphical models, undirected graphical models and hybrid graphical models.
    Specifically, directed graphical models are mainly based on hidden variables (i.e., variables not observed in the sample), 
    leveraging similarity of nodes or block structures, to generate the observed edges in a network; undirected graphical models are usually based on field structures, using the constraints of unary and pairwise potentials (e.g., the community label agreement between nearby nodes) to discover community; hybrid graphical models normally transform these two types of models into a unified factor graph to take advantages of both models for community detection.

    \textbf{Deep learning-based methods} aim to 
    identify community structures utilizing a new type of community-oriented network representation. It derives the new network representation through some learning strategies that map network data from the original input space to a low-dimensional feature space, with the advantages of low computational complexity and high capability for parallelization. 
    Depending on the learning strategies used,
    deep learning-based methods fall into four main categories: auto-encoder-based, generative adversarial network-based, graph convolutional network (GCN)-based, and integrating graph convolutional networks and undirected graphical models. Concretely, auto-encoder-based methods exploit unsupervised auto-encoder, which encodes a network into a low-dimensional representation in the latent space and reconstructs the network, along with its community structures, from the low-dimensional representation. Generative adversarial network-based methods adopt the idea of adversarial learning. They detect communities via an adversarial game between a generator and a discriminator. Graph convolutional  network-based methods extract communities by the propagation and aggregation of features on network topology. Hybrid graphical model-based methods integrate graphical convolutional networks and undirected graphical models by, for example, converting a Markov random field (MRF) layer into GCN, to take advantages of both models.

%% file: statistical_modeling.tex
\section{Community Detection with Probabilistic Graphical Model}
Probabilistic graphical models-based methods ordinarily detect communities through network modeling, i.e., employing graphical models to explain the generation process of networks.
In this section, we will focus on three general methods: 
directed graphical models, undirected graphical models, and hybrid graphical   models integrating directed and undirected graphs.
\subsection{Directed Graphical Models}
We will review the recent development of directed graphical models for community detection, including stochastic block model, topic model and matrix factorization.
These methods have solid theoretical basis and reasonably good performance, and have been broadly applied.

\subsubsection{Stochastic Block Model-based Methods}
Stochastic block model (SBM), an effective generative model of network block structures, adopts statistical modeling for community detection for the first time \cite{Holland1983Stoch}.
The method probabilistically assigns nodes in a network to different communities (block structures) using a node membership likelihood function, and then progressively infers the probabilities of node memberships by inferencing on the likelihood function to derive hidden communities in the network. 
Note that there are several SBM variants for community detection, but their core generation process is the same. The basic generation process can be divided into two steps: the first is to iteratively assign a community to each node in the network, and the second is to compute or update the probability of two nodes connected by an edge.

Taking a social network as an example, SBM can be used to capture a probabilistic generation process with the community distribution as a hidden variable. The communities can be reconstructed by maximizing a likelihood function of the node community membership. In this social network, the nodes are partitioned into $k$ disjoint communities with probability $\omega = {\left\{\omega_{1},\ldots,\omega_{k}\right\}}$. 
Assuming there are two nodes $v_i$ and $v_j$ belonging to two communities $\mathcal{C}_r$ and $\mathcal{C}_s$, represented by $c_{ir}$ and $c_{js}$. The probability that nodes $v_i$ and $v_j$ connected by an edge, i.e., $a_{ij}$ (0 or 1),
obeys a Bernoulli distribution with parameter $\pi_{rs}$. The use of a link probability between nodes in two communities makes the model flexible with various types of network structures\cite{lee2019review}. The network generation distribution can be defined as: 
\begin{equation}
\setlength\abovedisplayskip{8pt}
\setlength\belowdisplayskip{8pt}
    P(C|\omega)=\prod_{i=1}^n \operatorname{Multinomial}(c_i;1,\omega)=\prod_{i=1}^n \prod_{r=1}^k \omega_r^{c_{ir}},
\end{equation}
\begin{equation}
\setlength\belowdisplayskip{8pt}
\begin{split}
P(A|C,\pi)&=\prod_{i<j}^n
P(a_{ij}|c_i,c_j)= \prod_{i<j}^n \prod_{r,s}^k \operatorname{Bernoulli}(a_{ij}|\pi_{rs})^{c_{ir} c_{js}}\\
&= \prod_{i<j}^n \prod_{r,s}^k(\pi_{rs}^{a_{ij}}(1-\pi_{rs})^{(1-a_{ij})})^{c_{ir} c_{js}},
\end{split}
\end{equation}
where $C=(c_{1}, c_{2}, \ldots, c_{n})$ denotes the community assignment of nodes, and $\pi = (\pi_{rs})_{k\times k}$ represents the community connection probability matrix.
Then, the likelihood function 
can be described as:
\begin{equation}
\setlength\abovedisplayskip{8pt}
\setlength\belowdisplayskip{8pt}
\begin{split}
P&(A,C|\omega,\pi)= P(A|C, \pi)P(C|\omega)
\\= & \prod_{i=1}^n \prod_{r=1}^k \omega_r^{c_{ir}}
\times \prod_{i<j}^n \prod_{r,s}^k (\pi_{rs}^{a_{ij}}(1- \pi_{rs} )^{(1-a_{ij})})^{c_{ir}c_{js}}.  
\end{split}
\end{equation}
Based on the likelihood function, Gibbs sampling and the expectation-maximization (EM) algorithm can be used to obtain the model parameters, e.g., $\omega$ and $\pi$. 
Finally, we can derive the community partition with the interaction model between communities. The time complexity of the basic SBM is  $O({n^2}{k^2})$, and the process of community detection based on SBM with a Bernoulli distribution is shown in 
Appendix B \cite{Holland1983Stoch}. The generation process of the basic SBM may also employ a Poisson distribution \cite{snijders1997estimation}.

Later, Zhang \emph{et al}. \cite{2012Comparative} study the problem of hidden class inference in basic SBM. They present a comparative study between three inference approaches, i.e., heuristic spectral methods, mean field
approximation and belief propagation, on synthetic networks.


\textbf{Mixed membership SBMs.}
Since the basic SBM is only suitable under the assumption that a node belongs to only one community,
Airoldi \emph{et al}. \cite{Airoldi2008Mixed} propose a mixed membership stochastic blocks model (MMSB) that introduces mixed membership to the stochastic model so that one node may belong to multiple communities. MMSB allows communities to overlap on a directed network, where $a_{i\rightarrow j}$ indicates whether there is a link (arrow) from nodes $v_i$ to $v_j$. For each node $v_i$, $c_i$ obeys a Multinomial distribution. If $v_i \in \mathcal{C}_r$ and $v_j \in \mathcal{C}_s$, the community connection probability $\pi_{rs}$ follows a Beta distribution and $a_{i\rightarrow j} \sim \operatorname{Multinomial} (\omega_i)$, 
$a_{i\leftarrow j} \sim \operatorname{Multinomial} (\omega_j)$, where $\omega$ is the mixed membership parameter of nodes. The links between communities are represented by a Bernoulli distribution. 
The joint distribution of MMSB can be formulated as:
\begin{equation}
\setlength\abovedisplayskip{8pt}
\setlength\belowdisplayskip{8pt}
\begin{split}
    &P(A,\pi,\omega,a_{i \rightarrow j},a_{i \leftarrow j}|\alpha,\beta)=
    \prod_{i} P(\omega_i |\alpha) \times \\
    &P(\pi|\beta) \prod_{i,j} P(a_{i \rightarrow j}|\omega_i)P(a_{i \leftarrow j}|\omega_j ) P(a_{ij}|a_{i \rightarrow j},a_{i \leftarrow j},\pi). 
\end{split}    
\end{equation}
The time complexity of MMSB is $O({n^2}k)$, and the process of community detection based on MMSB is described in
Appendix B
\cite{Airoldi2008Mixed}, which assumes that the parameters are estimated by inference methods such as EM. 

The original MMSB is not good at handling diverse types of the information of nodes in community, e.g., the nodes may represent people who are connected one another based on different social relationships. To address this problem, Fan \emph{et al}.\cite{fan2016copula} propose a novel MMSB-based method, named Copula mixed membership stochastic block model (cMMSB) to introduce a Copula function into MMSB to model dependencies among nodes. 
Moreover, Miller \emph{et al.} \cite{2018Mixture} 
improve the inference of MMSB by introducing a Dirichlet process mixture to a mixture of finite mixtures (MFMs). Pal \emph{et al.} \cite{DBLP:conf/icassp/PalC19} propose a mixed membership degree-correct SBM and develop an inference method of the posterior distribution with Markov chain Monte Carlo (MCMC), to boost the embedding performance of MMSB. The degree-corrected SBM is widely used, which we will discuss next.



\begin{table*}[h!]
	\centering
	 \caption{\label{tab:sbm} Summary of SBM-based community detection, where "AD $k$" describes whether the approach can automatically determine the number of communities, i.e., Yes or No.} 
	\scalebox{0.99}{
	\begin{tabular}{p{1.5cm}<{\centering} p{3.5cm} p{7.8cm} c c}
			\toprule 
		{\bf Categories}      & {\bf Approaches} &  {\bf Sketches}  &  {\bf Overlapping}  &  {\bf AD $k$}\\
			\midrule
	
		 	Basic &SBM (1983) \cite{Holland1983Stoch}     & Propose a stochastic model for social network.   & No & No \\ 
			\hline
			 \multirow{6}{*}{MMSB} &  \multirow{1.8}{*} {MMSB (2008) \cite{Airoldi2008Mixed}}   & Extend blockmodels for relational data to ones that capture mixed membership latent relational structure. &\multirow{1.8}{*}{Yes} &\multirow{1.8}{*}{No}\\
			 \cline{2-5}
			 	{} &\multirow{1.8}{*} {cMMSB (2016)\cite{fan2016copula}}   & Combine an individual Copula function with MMSB with improved in capturing group interactions.
 &\multirow{1.8}{*}{Yes} & \multirow{1.8}{*} {No}\\
			 \cline{2-5}
			 	{} & \multirow{3}{*} {MMDCB (2019)\cite{DBLP:conf/icassp/PalC19}} & Propose a mixed membership degree-correct SBM and develop an inference method of the posterior distribution with Markov chain Monte Carlo (MCMC). &\multirow{3}{*} {Yes} &\multirow{3}{*} {No} \\
			 
			 \hline
			 \multirow{8}{*}{DCSBM} & \multirow{1.8}{*} {DCSBM (2012)\cite{zhao2012consistency}} & Introduce expected values to basic SBM to adapt 
			 multi-edges and self-edges contained in social networks. & \multirow{1.8}{*} {Yes}  &  \multirow{1.8}{*} {No} \\
			 \cline{2-5}
			 {} & \multirow{1.8}{*} {sparseDCSBM (2017)\cite{gulikers2017spectral}}  &Propose a spectral clustering algorithm with normalized adjacency matrix based on DCSBM. &\multirow{1.8}{*} {No} & \multirow{1.8}{*} {Yes}\\
			\cline{2-5}
			{} &\multirow{1.8}{*} {CMM (2018)\cite{chen2018convexified}}  &  Establish a convexified modularity maximization approach for estimating the hidden communities based on DCSBM.  &\multirow{1.8}{*} {No} & \multirow{1.8}{*} {No}\\
			\cline{2-5}
			{} & \multirow{1.8}{*} {NCV (2018)\cite{2014Network}} & 
			 Provide a network cross-validation approach to determine the hidden communities based on DCSBM. & \multirow{1.8}{*} {No}  &  \multirow{1.8}{*} {No} \\
			\hline
			 \multirow{15}{*}{DynSBM} &
			 \multirow{1.8}{*} {dMMSB (2009)\cite{fu2009dynamic}} & Propose a state space MMSB which can track
across time dynamic evolution. & \multirow{1.8}{*} {Yes} &\multirow{1.8}{*} {No} \\
\cline{2-5}
			 {} &\multirow{1.8}{*} {DynamicSBM (2010)\cite{Xing2008A}}  & Propose a novel Bayesian approach for network tomographic inference buliding on MMSB and apply in dynamic network. & \multirow{1.8}{*} {No} &  \multirow{1.8}{*} {No} \\
			 
			\cline{2-5}
			{} &\multirow{1.8}{*} {DSBM (2011)\cite{DBLP:journals/ml/YangCZGJ11}} &Capture the evolution of community by modeling the transition of community memberships for individual nodes. 
			&\multirow{1.8}{*} {No} &\multirow{1.8}{*} {No}\\
			\cline{2-5}
			{} &\multirow{1.8}{*} {DBTDP (2014)\cite{tang2014detecting}} & Propose a dynamic stochastic block model with temporal Dirichlet process for hidden community. &\multirow{1.8}{*} {Yes} &\multirow{1.8}{*} {No}\\
			\cline{2-5}
			{} &\multirow{1.8}{*} {SBTM (2015)\cite{xu2015stochastic}}  & 
		Provide a local search algorithm for the inference procedure of time evolution. & \multirow{1.8}{*} {No} &\multirow{1.8}{*} {No} \\
			\cline{2-5}
			{} &\multirow{1.8}{*} {dDCSBM (2016)\cite{wilson2016modeling}} & Propose a dynamic DCSBM to model and monitor dynamic networks that undergo a significant structural change. & \multirow{1.8}{*} {No} &\multirow{1.8}{*} {Yes}\\
			\cline{2-5}
			{} &\multirow{1.8}{*} {DPSBM (2019)\cite{wu2019dynamic}} & Establish a fully Bayesian generation model with 
			the heterogeneity of node degree. 
			& \multirow{1.8}{*} {No} &\multirow{1.8}{*} {Yes}\\
			\cline{2-5}
			{} &\multirow{3}{*} {SNR-DSBM/ER (2020)\cite{Bhattacharjee2020}}  &Focus on estimating the location of a single change point in a dynamic stochastic block model and take a least squares criterion function 
			for evaluating each point in time. &\multirow{3}{*} {No} & \multirow{3}{*} {Yes}\\
			\hline
			 \multirow{6}{*}{OSBM} &\multirow{1.8}{*} {OSBM (2011)\cite{latouche2011overlapping}}  &Provide a global and local variational technique for discovering community. &\multirow{1.8}{*} {Yes} & \multirow{1.8}{*} {No} \\
			 \cline{2-5}
			 {} &\multirow{1.8}{*} {K-LAFTER (2018)\cite{DBLP:conf/ijcai/AroraPASR18}} & Present a small-variance asymptotics based SBM for overlapping community detection. &\multirow{1.8}{*} {Yes} &\multirow{1.8}{*} {Yes}\\
			 \cline{2-5}
			 {} & \multirow{1.8}{*} {MNPAOCD (2020)\cite{jin2020modeling}} &  Optimize the inference process and expect parameters in proceeding. &\multirow{1.8}{*} {Yes} &\multirow{1.8}{*} {Yes}\\		
			 \hline
			 \multirow{1.8}{*}{LSBM} &\multirow{1.8}{*} {LMBP (2015)\cite{DBLP:conf/aaai/HeLJZ15}} & Combine heterogeneous distribution with SBM to  link community detection. & \multirow{1.8}{*} {Yes} & \multirow{1.8}{*} {Yes}\\
			 \hline
			 \multirow{1.8}{*} {GNNSBM} &\multirow{1.8}{*} {DGLRFM (2019)\cite{DBLP:conf/icml/MehtaCR19}}  &Design
			 a GNN-based overlapping SBM framework and can be adapted readily for other types of SBMs.  &\multirow{1.8}{*} {Yes} & \multirow{1.8}{*} {Yes}\\
			 \bottomrule
	\end{tabular}
	}
	\vspace{-0.3cm}
\end{table*}

\textbf{Degree-corrected SBMs.} Newman \emph{et al.} \cite{dcsbm} reason that the basic SBM divides nodes according to their degrees that are usually nonuniform distributed. To accommodate the possible broad degree distributions, they propose degree-corrected SBM (DCSBM), which introduces a degree parameter to every node to scale the edge probabilities and make expected degrees match observed degrees. 
The probability function of network $G$
is defined as follows:
\begin{equation}
\setlength\abovedisplayskip{8pt}
\setlength\belowdisplayskip{8pt}
\begin{split}
 P(G|\pi,c) &= \prod_{i<j} \frac{(\pi_{c_i c_j})^{a_{ij}}
}
 { a_{ij}!} \operatorname{exp}(-\pi_{c_ic_j})\\&
 \times \prod_{i} \frac{(\frac{1}{2}\pi_{c_ic_i})^{a_{ii}/2}}{ (a_{ii}/2)!} \operatorname{exp}(-\frac{1}{2}\pi_{c_ic_i}),\\
\end{split}
\end{equation}
where $\pi_{c_ic_j}$ is the expected value of the adjacency matrix element $a_{ij}$.

There are also several variants of DCSBM model. 
Gulikers \emph{et al}. \cite{gulikers2017spectral} propose an improved DCSBM which is feasible for sparse networks.
Some other distinct lines of effort are to improve and extend DCSBM by model inference. 
Chen \emph{et al}. \cite{chen2018convexified} present a convexified modularity maximization approach for estimating hidden communities under DCSBM. 
Chen \emph{et al}. \cite{2014Network} provide a  network cross-validation (NCV) approach based on a block-wise node-pair splitting to determine the hidden communities for DCSBM.

\textbf{Dynamic SBMs.} Different from the above methods, analyzing dynamic networks based on SBM is also a relatively active field.
Yang \emph{et al}. \cite{DBLP:journals/ml/YangCZGJ11} suggest a dynamic stochastic block model named DSBM, which progressively updates the probabilistic model to find community in large dynamic sparse network. 
Specifically, DSBM uses the distribution of model parameter instead of the most likely values for the model parameters in prediction, and provides an offline inference and an online inference to estimate the parameters.
DSBM assumes that nodes in a dynamic network remain unchanged.
Letting $C_{T} =\{C^{(1)}, C^{(2)},...,C^{(T)}\}$  be the collection of community assignments of all nodes over $T$ discrete time steps, and the likelihood function of the model is as follows:
\begin{equation}
\setlength\abovedisplayskip{8pt}
\setlength\belowdisplayskip{8pt}
    \begin{split}
       &P(W^{(T)},C^{(T)}|\omega,\pi,A) =\\ & \prod_{t=1}^TP(W^{(t)}|C^{(t)},\pi)
        \prod_{t=2}^TP(C^{(t)}|C^{(t-1)},A)P(C^{(1)}|\omega),
    \end{split}
\end{equation}
where $W^{(t)}$ and $C^{(t)}$ denote the snapshot of a network and the community assignments of nodes at a given time step $t$. The time complexity of DSBM is $O(nT+mT+k^2 T)$, and its generation process is illustrated in Appendix B \cite{DBLP:journals/ml/YangCZGJ11},





Following DSBM, Tang \emph{et al}.\cite{tang2014detecting} introduce the Dirichlet process to SBM, which can find the optimal number of communities of evolution, and in turn, alleviate the problem of fixed community number of dynamic social networks. Xu \emph{et al}. \cite{xu2014dynamic} propose a new approach named stochastic block transition model (SBTM) that includes two hidden Markov assumptions for dynamic network. Wu \emph{et al}. \cite{DBLP:conf/dasfaa/WuJWLWW19} propose a full Bayesian generation model, which incorporates the heterogeneity of the degrees of nodes to model dynamic complex networks. Bhattacharjee \emph{et al}. \cite{Bhattacharjee2020} optimize SBM with change-point estimation in dynamic social networks. Inspired by the success of MMSB, some dynamic SBMs based on MMSB have been proposed. Xing \emph{et al}.\cite{Xing2008A} propose a variant MMSB model for dynamic networks. Fu \emph{et al}. \cite{fu2009dynamic} design a state space mixed membership stochastic block model with crossing time. 
Besides, there also exists DCSBM-based approaches. Wilson \emph{et al}. \cite{wilson2016modeling} suggest a dynamic version of DCSBM to model and monitor dynamic networks that undergo a significant structural change.

\textbf{Others.} In addition to the above methods, there are several other extensions to the basic SBM where communities can overlap, as summarized in Table 2. 
For instance, OSBM represents the SBM that are designed to find overlapping communities and LSBM represents the SBM that are extended to find link community. Specifically, Latouche \emph{et al}. introduce OSBM with global and local variational technology. Jin \emph{et al}. \cite{jin2020modeling} provide a stochastic model to accommodate the relative importance and the expected degree of every node in each community, 
and improve the inference technique that it uses.

Link communities are often more informative and intuitive than node communities, because links usually have unique identities, whereas nodes may have multiple roles. For instance, in a social network, most individuals belong to multiple communities such as families or friends, while the link between two individuals often exists for a dominant reason, which may represent family ties or friendship. 
Furthermore, multiple links connecting to a node may belong to distinct link communities, so that the node can be assigned to multiple communities of links. He \emph{et al}. \cite{DBLP:conf/aaai/HeLJZ15} combine heterogeneous distributions (e.g., power law distribution) of community sizes
with SBM for link community detection. They suggest a stochastic model for link community and extend the model by introducing a scheme of interactive bipartition. 
Besides the above models, Mehta \emph{et al}. \cite{DBLP:conf/icml/MehtaCR19} introduce graph neural network into SBM, which integrate deep learning and SBM for the first time.


\subsubsection{Topic Model-based Methods}
Topic model, such as Latent Dirichlet Allocation (LDA)\cite{blei2003latent}, is a statistical model capable of modeling hidden topics behind texts in natural language processing. LDA models topics by employing latent variables, which have attracted significant interests and have been widely used in detecting communities. Topic models can be grouped into two categories: one models network structures as documents and the other models attributes of network, such as user interests, to detect communities.

\textbf{Modeling network structures as documents.}  We take LDA as an example to describe the principle of the methods in the first category. To be specific, a method in this group first assumes that each node in a network may belong to multiple communities, and thus the communities are regarded as "topics" while the nodes are taken as "documents". 
It then selects several initial communities, and iteratively updates the communities according to the topology of the network to obtain resulting communities.
Among the existing methods, a representative model is SSN-LDA \cite{DBLP:conf/isi/ZhangQGFY07}, 
which is a LDA-based hierarchical Bayesian algorithm on link networks where communities are modeled as latent variables. Nodes in such a social network are regarded as social actors and edges as social interaction. Social interaction profile (SIP) of each social actor, consisted of a set of neighbors and weights, is used to characterize the actor. Specifically, in SSN-LDA, a social network is viewed as a corpus, where social interaction profiles are regarded as documents and the occurrence of social interaction is deemed as words.
The nodes are modeled as a corpus by SSN-LDA, which mines communities on transformed corpus, and this problem is equivalent to topic detection on corpus utilizing LDA. The generation process of SSN-LDA for one social interaction profile $(SIP_i)$ is clarified in Appendix C \cite{DBLP:conf/isi/ZhangQGFY07}, and the joint distribution is written as:
\begin{equation}
\setlength\abovedisplayskip{8pt}
\setlength\belowdisplayskip{8pt}
    \begin{split}
     P(a_{i},c_{i},\vec{\theta_{i}},\vec{\phi}|\vec{\alpha}, \vec{\beta})=\prod_{j=1}^{N_i}P(a_{ij}|\vec{\phi_{c_{i}}})P(c_{i}|\vec{\theta_{i}}) P(\vec{\theta_{i}}|\vec{\alpha}) P(\vec{\phi}|\vec{\beta}), 
    \end{split}
\end{equation}
where $\vec{\phi}$ is the mixture component of community $c_i$, $N_i$ is the number of social interactions in a social interaction profile ($SIP_i$), 
$\vec{\theta_{i}}$ is the community mixture proportion for $SIP_i$, and $\vec{\alpha}$ and $\vec{\beta}$ are the Dirichlet prior distribution hyper-parameters that are known. The time complexity of  SSN-LDA is $O(mk)$.

\textbf{Using social network attributes.} Numerous topic models utilize attributes of social network, e.g., user interests, to discover community.
Yin \emph{et al}. \cite{Yin2012Latent} propose to integrate community detection and topic model, which gives rise to latent community topic analysis (LCTA). Their method divides the sampling process into user node and link samplings. The process is to sample all network connections after sampling a user node, and exploit the sampling results of these two stages as the sampling result of user node. LCTA assigns community membership attributes to each user node and link. After the sampling process, user nodes can be assigned to communities based on community membership. The advantage is that the two-stage sampling process forms a sampling area with user nodes as the core, which can simulate the semantic influence of user nodes on surrounding links. The disadvantage is that LCTA does not consider the link relationship of social network when assuming the degree of community membership, which may disconnect individual communities. Further, Cha \emph{et al}. \cite{DBLP:conf/sigir/ChaC12} design a tree relationship model according to the topic information of followers in social network, use hierarchical LDA to model the text information in tree relationship model, and propose HLDA for semantic social network analysis.

A method of \textbf{combining topic model with Bayesian model} is proposed recently by Xu \emph{et al}. \cite{xu2012model}. They define a joint probability distribution 
for all possible attributed networks. For a given attributed social network to be clustered, 
the model assigns a probability to each possible clustering of nodes. Therefore, the clustering problem can be transferred to the problem of finding the clusters that have the highest probability. The algorithm for clustering attribute communities is shown in Appendix C \cite{xu2012model}.
The Bayesian probabilistic model for clustering attributed networks is as follows:
\begin{equation}
\setlength\abovedisplayskip{8pt}
\setlength\belowdisplayskip{8pt}
    \begin{split}
        &P(\alpha, \theta, \phi, A, X, C|\varepsilon, \lambda, \mu, \nu)
        \\&=P(\alpha|\varepsilon) P(\theta|\lambda) P(\phi|\mu, \nu) P(C|\alpha) P(A|C,\phi) P(X|C,\theta),
    \end{split}
\end{equation}
where $\alpha$ denotes the probability of the nodes belonging to different communities, $\theta$ represents the attribute probability distribution of nodes, $\phi$ denotes the edge occurrence probability between communities, $\varepsilon$ and $\lambda$ are the Dirichlet prior distribution hyper-parameters, and $\mu$ and $\nu$  are the Beta prior distribution hyper-parameters.

Later, He \emph{et al}. \cite{DBLP:conf/aaai/HeFJWZ17} introduce a generative model for simultaneously identifying communities and deriving their semantic description. They combine a nested EM algorithm with belief propagation, and explore the hidden correlation between the two parts to improve resulting communities and description. 
Jin \emph{et al}. \cite{jin2019detecting} 
observe that the attributes usually embody a hierarchical semantic structure. To handle this, they propose a novel Bayesian model named BTLSC, which distinguishes words from background and general from specialized topics. 

Unlike traditional topic models that assume that the topics of social network are independent, \textbf{topic embedding methods} focus on describing correlations between topics by embedding words and topics into topic models. He \emph{et al}. \cite{DBLP:conf/kdd/HeHBHX17} present a topic embedding model that combines distributed representation learning with topic correlation modeling. 
Jin \emph{et al}. \cite{DBLP:conf/www/JinHJ0HSH19} 
develop a novel topic embedding model named community-enhanced topic embedding (CeTe), which combines topic documents and network structures to detect communities. CeTe consists of three components: a document component for describing topics, a topological component for representing network communities, and a probabilistic transition mechanism connecting the first two parts. Specifically, CeTe uses a DCSBM to describe the sub-component of network community, where communities obey a Dirichlet distribution, and topics obey a Uniform distribution. For each document, the community assignment is drawn from a Multinomial distribution, whereas the link between two documents obeys a Bernoulli distribution. For each word, CeTe draws topic distribution following a Multinomial distribution.



\subsubsection{Matrix Factorization-based Methods}
Non-negative matrix factorization (NMF) \cite{DBLP:conf/nips/LeeS00} is another directed graphical model for community detection. 
Specifically, 
the NMF-based methods assume there are $k$ communities in a network, and deem the adjacency matrix $A=(a_{ij})_{n\times n}\in\mathbb{R}_{+}$ as a non-negative matrix to be decomposed, where $a_{ij}$ denotes the likelihood if there is a connection between nodes $v_i$ and $v_j$. We define
$\mathcal{W}=(w_{ir})_{n\times k}\in \mathbb{R}_{+}$ and $\mathcal{H}=(h_{jr})_{n\times k}\in\mathbb{R}_{+}$, whose elements $w_{ir}$ and $h_{jr}$ represent the likelihoods that $v_{i}$ generates an out-edge, i.e., an edge starting from $v_{i}$, and $v_{j}$ generates an in-edge, i.e., an edge ending at $v_{j}$, belonging to the $r$-th community. Then, the likelihood that 
nodes $v_i$ and $v_j$ are connected
can be described as:
\begin{equation}
\setlength\abovedisplayskip{8pt}
\setlength\belowdisplayskip{8pt}
\widetilde{a_{ij}} = \sum_{r=1}^{k}w_{ir}h_{jr}^{T}.
\end{equation}
As a result, the community detection problem can be represented as $\widetilde{A} = \mathcal{W}\mathcal{H}^{T}$.
In general, there are two classic loss functions to evaluate the performance of NMF. One is the square of the Frobenius norm of the difference between $A$ and $\widetilde{A}$\cite{wang2008clustering},
and the other is the KL-divergence 
that measures their difference.

\begin{table*}
\caption{Summary of NMF-based community detection, where "AD $k$" describes whether the approach can automatically determine the number of communities, i.e., Yes or No.}
\begin{center} 
\begin{tabular}{m{2.1cm}<{\centering} m{2.75cm} m{4.55cm} m{3.9cm} m{1.3cm}<{\centering} m{1.0cm}<{\centering}}
\toprule
\bf Categories & \bf Approaches  & \bf Sketches & \bf Objective Functions & \bf Overlapping & \bf AD $k$\\

\midrule 
\multirow{7}{*}{Basic} & SymNMF (2012) \cite{kuang2012symmetric} & Develop a symmetric NMF framework based on Newton-like for graph clustering. & $\left \|A - BB^{T}\right \|_{F}^{2}$ & \multirow{1}{*}{No} & \multirow{1}{*}{No} \\

\cline{2-6} 
& PCSNMF (2015) \cite{shi2015community}  & Present a symmetric NMF method with pairwise constraints generated from the ground-truth community information.& $\left \|A - BB^{T}\right|_{F}^{2}$+

$\alpha[\operatorname{Tr}(B^{T}\mathcal{M}B\mathcal{Q})+\operatorname{Tr}(B^{T}\mathcal{P}B)]$
& \multirow{1}{*}{No} & \multirow{1}{*}{No} \\

\cline{2-6} 
& NSED (2017) \cite{sun2017non} & Pose a non-negative symmetric encoder and decoder approach to obtain a better network representation. & $\left \|A - \mathcal{W}\mathcal{H}\right \|_{F}^{2}+\left \|\mathcal{H}- \mathcal{W}^{T}A \right \|_{F}^{2}$ & \multirow{1}{*}{No} & \multirow{1}{*}{No} 
\\

 \hline 
\multirow{5}{*}{Overlapping} & SNMF, ANMF,

JNMF (2011) \cite{wang2011community} & Apply NMF to community detection firstly. & $\left \|A - BB^{T}\right \|_{F}^{2}$ & Yes  & No \\

\cline{2-6} 
& BIGCLAM (2013) \cite{Yang2013} & Present a cluster affiliation model for overlapping, hierarchically nested community detection in large scale networks.
& $\sum_{e_{ij}\in E}\log(1-\exp(-b_{i}b_{j}^{T}))$

$-\sum_{e_{ij}\notin E}b_{i}b_{j}^{T}$  & \multirow{1}{*}{Yes} & \multirow{1}{*}{Yes}
 \\
 



\hline 
\multirow{4}{*}{Attribute}  & NMTF (2015) \cite{pei2015nonnegative} &  Develop a NMF clustering framework combining nodes' relations and users' contents to detect community structure. & $\left \|M_{u-u} - \mathcal{U}H_{1}\mathcal{U}^{T}\right \|_{F}^{2}$

$+\left \|M_{t-f} - \mathcal{V}H_{2}\mathcal{N}^{T}\right \|_{F}^{2}$

$+\left \|M_{u-f} - \mathcal{U}H_{3}\mathcal{N}^{T}\right \|_{F}^{2}$ & \multirow{1}{*}{No} & \multirow{1}{*}{No} 
\\

\cline{2-6} 
& SCI (2016) \cite{Wang2016}  & Propose a semantic community identification method, which can annotate semantic as well as detect community. & $\left \| B-XS \right \|_{F}^{2}$

$+\alpha \sum_{r=1}^{k}\left \|S\left ( :,r \right )  \right \|_{1}^{2}$

$+\beta \left \| A-BB^{T} \right \|_{F}^{2}$ & \multirow{1}{*}{No} & \multirow{1}{*}{No} 
\\

\hline 
\multirow{4}{*}{Dynamic}  & DBNMF (2016) \cite{DBLP:journals/kbs/WangJHJPG16} &  Present a Bayesian probabilistic model based on NMF to identify overlapping communities in temporal networks. &$-\log P(V_{t}|B_{t})$

$-\log P(B_{t}|B^{''}_{t-1},\alpha ) $

$-\log P(B_{t}|\beta _{t})-\log P(\beta _{i})$ & \multirow{1}{*}{Yes} & \multirow{1}{*}{Yes}  \\

\cline{2-6} 
& sE-NMF (2017) \cite{Ma2017} & Develop a semi-supervised evolutionary NMF framework for dynamic community detection via prior information. & $\left \|A_{t}-\tilde{B_{t}}\tilde{B_{t}}'\right \|_{F}^{2}$ & \multirow{1}{*}{No} & \multirow{1}{*}{No}  \\

\hline
\multirow{1.2}{*}{Semi-supervised} & USSF (2015) \cite{DBLP:journals/tcyb/YangCJWM15} &  Present a unified semi-supervised community detection algorithm based on combination of prior and topology information aimed NMF.  &
$\mathcal{L}_{\alpha}(A,B)+\lambda \mathcal{R}_{\beta}(O,B)$ & \multirow{1}{*}{No} & \multirow{1}{*}{No}  \\

 \bottomrule
\end{tabular}
\end{center} 
\vspace{-0.3cm}
\end{table*}
Furthermore, for a undirected network with $A$ being symmetric, the non-negative factorization matrices $\mathcal{W}$ and $\mathcal{H}$ should be equal. In this paper, we use $B$ to represent these matrices, 
and the loss function can be written as:
\begin{equation}
J = \min_{B\geq 0}\left \|A - BB^{T}\right \|_{F}^{2}.
\end{equation}
NMF is initially used to identify non-overlapped community. Since it is easily extendable, NMF has been adopted to solve other types of community detection problems, such as overlapping, attributed, dynamic and semi-supervised, as summarized in Table 3. 

\textbf{Basic NMF.} Kuang \emph{et al}. \cite{kuang2012symmetric} propose a general approach, which inherits the advantages of NMF by enforcing non-negativity on the clustering assignment matrix, for graph clustering. Shi \emph{et al}. \cite{shi2015community} present a novel pairwise constrained non-negative symmetric matrix factorization (PCSNMF) method, which imposes pairwise constraints generated from ground-truth community information, to improve the performance of community detection. Sun \emph{et al}. \cite{sun2017non} design a non-negative symmetric encoder-decoder approach to derive a better latent representation to improve community detection. Unlike other NMF-based methods that merely pay attention to the loss of the decoder, they combine the loss of the decoder and encoder to construct a unified loss function, so that the community membership of each node obtained is clearer and more explanatory.

\textbf{Overlapping NMF.} Overlapping community detection is another active research topic due to the overlapping and nesting properties of real-world networks. Wang \emph{et al}. \cite{wang2011community} develop a NMF framework to identify non-overlapping and overlapping community structure, and give a symmetric NMF formula for undirected networks. Moreover, they clarify the methods on asymmetric NMF and joint NMF, where the former is capable of identifying  community structures in directed networks, while the latter is more suitable for compound 
networks (e.g., an automatic movie recommendation system which contains three networks: user network, movie network and user-movie network). Yang \emph{et al}. \cite{Yang2013} present a cluster affiliation model BIGCLAM to detect densely overlapping, hierarchically nested, and non-overlapping communities in a massive network.
Specifically, BIGCLAM first builds on communities based on community affiliation of nodes, i.e., each node has an affiliation strength to each community via assigning node-community pair a non-negative latent factor, and then combines the NMF with block stochastic gradient descent, so as to estimate the non-negative latent factors to detect communities in large networks. The loss function of the model is defined as:
\begin{equation}
J=\sum_{e_{ij}\in E}\log(1-\exp(-b_{i}b_{j}^{T}))- \sum_{e_{ij}\notin E}b_{i}b_{j}^{T}.    
\end{equation}
The time complexity of BIGCLAM is $O(n)$.

\textbf{Attribute NMF.} Recently, it has attracted a substantial amount of interest to the semantic information of community structure utilizing NMF, i.e., delineating the corresponding community semantic information while identifying community structure \cite{pei2015nonnegative,Wang2016,guo2018cfond}.
In particular, Pei \emph{et al}. \cite{pei2015nonnegative} combine social relations and content of users to detect communities via a non-negative matrix tri-factorization (NMTF)-based clustering 
with three types of graph regularization. 
However, the above method merely exploits network topology and content information to discover communities,
without considering how to utilize the mined contents, i.e., semantic information, to explain the meaning of communities. To address this issue, wang \emph{et al}. \cite{Wang2016} propose a semantic community identification called SCI, which integrates the community membership matrix denoting network topology and community attribute matrix representing semantic information. 
The loss function of the model is defined as: 
\begin{align}
\setlength\abovedisplayskip{8pt}
\setlength\belowdisplayskip{8pt}
\begin{split}
\min_{B\geq 0, S\geq 0}J =& \left \| B-XS \right \|_{F}^{2}+\alpha \sum_{r=1}^{k}\left \|S\left ( :,r \right )  \right \|_{1}^{2} \\
&+\beta \left \| A-BB^{T} \right \|_{F}^{2},
\end{split}
\end{align}
where $S$ represents attribute community matrix, $\alpha$ is a trade-off hyper-parameter between the first error and the second sparsity term, and $\beta$ is a positive parameter for setting the proportion of the contribution of network topology. The time complexity of SCI is $O((nqk + n^{2}k))$, where $q$ is the dimension of node attribute.

\textbf{Dynamic and semi-supervised NMF.}
It deserves further attention that, during recent years, several investigators have extended NMF in the field of dynamic and semi-supervised community detection, and achieved encouraging results. For \textbf{dynamic} community detection, Wang \emph{et al}. \cite{DBLP:journals/kbs/WangJHJPG16} utilize a Bayesian model based on NMF to identify overlapping communities on temporal networks, and automatically derive the number of communities in each snapshot network based on automatic relevance determination. The loss function is as follows:
\begin{align}
\begin{split}
J_{t} =&-\log P(V_{t}|B_{t})-\log P(B_{t}|B^{''}_{t-1},\alpha ) \\
&-\log P(B_{t}|\beta _{t})-\log P(\beta _{i}), \\
\end{split}
\end{align}
where $V_{t}$ is a snapshot of a temporal network, $B_{t}$ is the non-negative matrix obtained via $V_{t}$, and $B^{''}_{t-1}$ is the new $B_{t-1}$ which has been adjusted according to the node distribution of $B_{t}$.
$\beta_{(\cdot)}$ is a parameter from a half normal distribution, and $\alpha$ is a parameter to balance the clustering results of the current and previous snapshot networks. 
Later, Ma \emph{et al}.\cite{Ma2017} show that NMF can be applied to dynamic community detection via clarifying the equivalence relationship among evolutionary spectral clustering, evolutionary NMF and optimization of evolutionary modularity density. Therefore, they employ the above equivalence relationship to develop a semi-supervised evolutionary NMF method, named sE-NMF, to integrate the prior information to detect communities in dynamic temporal networks.

\begin{table*}
\setlength{\abovecaptionskip}{0pt}
\setlength{\belowcaptionskip}{0pt}
\caption{Summary of MRF-based community detection.}
\begin{center} 
\begin{tabular}{m{2.7cm}<{\centering} m{2.8cm} m{6cm} m{4.4cm}}
\toprule
\bf Categories & \bf Approaches & \bf Sketches & \bf Object Functions \\

\midrule 
\multirow{7}{*}{Topology} & \multirow{1.8}{*}{NetMRF (2018) \cite{He2018}} & \multirow{1.8}{*}{Apply MRF to community detection firstly.} & $\sum\limits_{v_i\neq v_j}[-(-1)^{\delta (c_i, c_j)}(\frac{d_id_j}{2m} - a_{ij})]$  \\
 
\cline{2-4} 
&\multirow{1}{*}{GMRF (2019) \cite{JinD.2019}} & Optimize network embedding and develop a general MRF framework by incorporating network embedding into MRF to better detect community structure.
& $\sum\limits_{v_i}\Theta _{i}(c_i) + \sum\limits_{e_{ij} \in E} \Theta_{ij}(c_i, c_j)$ \\

\cline{2-4} 
&\multirow{1.8}{*}{ModMRF (2020) \cite{DBLP:journals/ijon/JinZSHFCLM20}} &Propose a MRF method formalizing modularity as the energy function for community detection. 
& $\sum\limits_{v_i, v_j \in V} -(a_{ij} - \frac{d_id_j}{2m})\delta (c_i, c_j)$\\

\hline 
\multirow{1.9}{*}{Topology \& attribute} & \multirow{1.9}{*}{ attrMRF (2019) \cite{HeDongxiao2019}} & Present a model integrating LDA into MRF to form an end-to-end learning system for community detection. & $
\sum\limits_{v_i \neq v_j} \Theta _{ij}(c_i, c_j; a_{ij}) $

$- \sum\limits_{r=1}^{n}\frac{1}{\beta _{1}}\ln f_{\theta_{r}}- \sum\limits_{p=1}^{q}\frac{1}{\beta _{1}}\ln f_{\phi_{p}}  $\\

\hline 
\multirow{4.5}{*}{Combining GNN}  & \multirow{1}{*}{MRFasGCN (2019) \cite{6}}  & Design a new approach based on the combination of GCN and MRF for semi-supervised community detection.  &$
-\sum\limits_{i=1}^{n'}\sum\limits_{j=1}^{k}Y_{ij}\operatorname{ln}Z_{ij}$\\

\cline{2-4}
&\multirow{1}{*}{GMNN (2019) \cite{7}} & Propose a new  approach combining the advantages of both statistical relational learning and graph neural network for semi-supervised node classification.  & $E_{q_{\theta}}(y_{U}| x_{V})[\log p_{\phi}(y_{L},y_{U}|x_{V})$

$- \log q_{\theta}(y_{U} | x_{V})] $     
\\

\bottomrule
\end{tabular}
\end{center}
\vspace{-0.5cm}
\end{table*}

For \textbf{semi-supervised} community detection,  Yang \emph{et al}.\cite{DBLP:journals/tcyb/YangCJWM15} put forward a unified semi-supervised algorithm by combining prior information and topology information aimed at two non-negative matrices generated by NMF. Moreover, with the must-link prior information (i.e., the prior information that a node pair composed of two nodes must belong to the same community \cite{Ma2010}), they add a graph regularization item as a penalty function to the loss function to minimize the difference between nodes in the same community, thereby improving the performance of community detection. The loss function is defined as:
\begin{equation}
J(B|A,O)=\mathcal{L}_{\alpha}(A,B)+\lambda \mathcal{R}_{\beta}(O,B),   \end{equation}
where $O$ denotes the matrix of prior information. $\mathcal{L}_{\alpha}(A,B)$ is the loss function of NMF, where ${\alpha} \in $ \{LSE, KL, SYM, MOD, ADJ, LAP, NLAP\} is the parameter to measure similarity. $\lambda \mathcal{R}_{\beta}(O,B)$ is a graph regulation term, where $\lambda$ is the tradeoff parameter between the loss function and graph regulation term, and $\beta \in \{ LSE, KL\} $ is the specific graph regulation term.


\subsubsection{Summary for Direct Graphical Models}
The direct graphical models typically transform the community detection problem into a Bayesian inference problem based on the observed network data and then optimize model parameters utilizing a maximum likelihood function or a maximum of a posteriori to obtain hidden variables of the model to discover network community structures. However, these methods often ignore the diversity of community patterns (e.g., community structure with homophily or heterophily) in the real world, and the network topology used is often noisy and sparse, limiting the performance of community detection.

\subsection{Undirected Graphical Models}
To the best of our knowledge, the existing studies of undirected graphical models for community detection mainly exploit Markov random field (MRF) \cite{DBLP:journals/ftcgv/NowozinL11}. MRF, a kind of random field, 
has enjoyed much success covering a variety of applications, such as computer vision and image processing. Particularly, we are interested in its applications to community detection. The MRF-based methods can be grouped into three categories (as summarized in Table 4): the first is the modeling based on MRF that detects community relation based on network topology, the second exploits the information of semantic attributes, and the third combines MRF with graph neural network (GNN). 

\textbf{Topology MRF.} He \emph{et al}. \cite{He2018} first apply MRF to network analysis where data are organized on networks with irregular structures, and propose a network-specific MRF approach, namely NetMRF, for community detection. This method effectively encodes the structural properties of an irregular network in an energy function so that the minimization of the energy function gives rise to the best community structures. The energy function can be represented as the sum of pairwise potential functions, written as follows:
\begin{align}
\begin{split} \label{pairwise}
E(C;A) = & \sum_{v_i\neq v_j}\Theta _{ij}(c_i, c_j; a_{ij}) \\ 
=& \sum_{v_i\neq v_j}[-(-1)^{\delta (c_i, c_j)}(\frac{d_id_j}{2m} - a_{ij})],  
\end{split}
\end{align}
where $\delta$ is the probability of nodes $v_i$ and $v_j$ falling into the same community partition, $d_{i}$ is the degree of node $v_i$, and $m$ is the numbers of edges. 
According to \cite{DBLP:journals/ftcgv/NowozinL11}, the smaller the function, the better the community partition. Further, Jin \emph{et al}. \cite{DBLP:journals/ijon/JinZSHFCLM20} formalize the modularity function as a statistical model and propose a novel MRF method for community detection. This method redefines the energy function via the approach of modularity representation, and leverages the max-sum belief propagation (BP) to infer model parameters to improve the performance. The energy function is represented as follows:
\begin{equation}
\setlength\abovedisplayskip{8pt}
\setlength\belowdisplayskip{8pt}
\begin{split}
E(C;A) 
= \sum_{v_i, v_j \in V} -(a_{ij} - \frac{d_id_j}{2m})\delta (c_i, c_j).
\end{split}
\end{equation}
The time complexity is $O(m)$.

Moreover, to overcome the issue of losing vital structural information between nodes after network embedding, Jin \emph{et al}. \cite{JinD.2019} propose a general MRF method to incorporate coupling relationship between pairs of nodes 
in network embedding to better detect community. In this method, the energy function is composed of two components: a set of unary potentials that make the network embedding to play a dominant role and a set of pairwise potentials that utilize constraints on node pairs to fine-tune unary potentials. Formally, the complete energy function can be defined as:
\begin{equation}
\setlength\abovedisplayskip{8pt}
\setlength\belowdisplayskip{8pt}
E(C;A) = \sum_{v_i}\Theta _{i}(c_i) + \sum_{e_{ij} \in E} \Theta_{ij}(c_i, c_j),
\end{equation}
where $\Theta _{i}$ and $\Theta_{ij}$ are the unary potential function and pairwise potential function respectively.

\textbf{Topology \& attribute MRF.} The combination of MRF and node semantic models (e.g., a topic model) have been a recent research focus. However, methods that directly integrate MRF with node semantic models
cannot in general achieve satisfactory results. It is mainly because the parameters of the two models cannot be adjusted to support each other, making it difficult to combine the advantages of the two methods. He \emph{et al}. \cite{HeDongxiao2019} propose a new model, named attrMRF, to integrate LDA \cite{blei2003latent} and MRF to form an end-to-end learning system to train the parameters jointly.
Concretely, attrMRF first transforms LDA and MRF into a unified factor graph, realizing the effective integration of directed graphic model (i.e., LDA) and undirected graphic model (i.e., MRF).
Then it adopts a backpropagation (BP)
algorithm to train the parameters simultaneously, resulting in an end-to-end learning of the two models.
The global energy function of this model is represented as:
\begin{equation}
\setlength\abovedisplayskip{8pt}
\setlength\belowdisplayskip{8pt}
\begin{split}\label{factor graph}
E(Z,C;A,X,\alpha, \beta ) &= \sum_{v_i \neq v_j} \Theta _{ij}(c_i, c_j; a_{ij}) \\
& - \sum_{r=1}^{n}\frac{1}{\beta _{1}}\ln f_{\theta_{r}}- \sum_{p=1}^{q}\frac{1}{\beta _{1}}\ln f_{\phi_{p}}, \end{split}    
\end{equation}
where $\sum_{v_i \neq v_j} \Theta _{ij}(c_i, c_j; a_{ij})$ denotes the global energy potential of MRF as defined in \eqref{pairwise}, $\beta_1$ is a temperature coefficient, $f_{\theta_{r}}$ and $f_{\phi_{p}}$ are the intermediate results generated by LDA joint probability distribution.
Besides attrMRF, there are also several approaches incorporating probabilistic graphical models into deep learning, such as MRFasGCN \cite{6} and GMNN \cite{7}, which will be covered in detail later.

The undirected graphical models mainly use the characteristics of MRF, i.e., unary and pairwise potentials for irregular network structure and node attribute, to identify community structures. Nevertheless, different energy functions need to be fine-tuned.

\subsection{Integrating Directed and Undirected Models}
Directed and undirected graphical models have also been integrated to detect communities in complex networks. This type of integration  is typically implemented via a factor graph model.
A factor graph \cite{DBLP:journals/pami/ZengCL13} is a tuple $\left (V, \mathcal{F}, \varepsilon \right )$ consisting a set $V$ of variable nodes, a set $F$ of factor nodes, and a set $\varepsilon \subseteq V\times \mathcal{F} $ of edges each of which connects a variable node and a factor node. Taking MRF as an example, the joint probability distribution of a factor graph is described as:
\begin{equation}
\setlength\abovedisplayskip{8pt}
\setlength\belowdisplayskip{8pt}
p(y) = \frac{1}{\mathbb{Z}}\prod_{F\in \mathcal{F}}\psi_{F}(y_{N(F)}),  
\end{equation}
where $\mathbb{Z} = \sum_{y \in Y}\prod_{F\in \mathcal{F}}\psi_{F}(y_{N(F)})$ denotes a normalization factor, and $N(F)=\left \{ v_{i} \in V:(v_{i},F) \in \varepsilon  \right \}$ is the set of variable nodes adjacent to factor node $F$. 

Yang \emph{et al}. \cite{yang2011social} first propose an instantiation model based on factor graph, which incorporates three layers, bottom layer (observed nodes), middle layer (hidden vector) and top layer (latent variables for communities).
It utilizes node-feature and edge-feature functions to mine dependencies between bottom
and top layer nodes to represent corresponding communities, 
so as to better detect communities. Further, Jia \emph{et al}. \cite{jia2014novel} apply factor graph model to ego-center network (a kind of representation of human social networks,
which is used to represent the network between an individual and others that the ego has a social relationship with \cite{passarella2012ego}), and propose an ego-centered method to analyze social academic influence on co-author networks. This method model the ego-centered community detection in a unified factor graph, employing a parameter learning algorithm to estimate the topic-level social influence, the social relationship strength between these nodes and community structures to detect ego-community structures.

These methods merely identify the structure of communities and ignore the semantic information of community, which is much critical for understanding the meaning of community structure. He \emph{et al}. \cite{HeDongxiao2019} employ a factor model to overcome the deficiency that directed graphical model (i.e., LDA) and undirected graphical model (i.e., MRF) are
insufficient to integrate due to parameter sharing and joint training and to make the discovered community structure semantically interpretable. The joint probability distribution of MRF and LDA formulated in factor graph is written as:
\begin{equation}
P(Z,C;A,X,\alpha,\beta)=\frac{1}{\mathbb{Z}}\prod_{r=1}^{n}f_{\theta _{r}}\prod_{p=1}^{q}f_{\phi_{p}}\prod_{v_i\neq v_j}f_{\gamma_{ij}},
\end{equation}
where $\mathbb{Z}$ denotes normalization term, $f_{\theta _{r}}$ and $f_{\phi _{p}}$ are defined in \eqref{factor graph},
and $f_{\gamma_{ij}}$ is the pairwise potential of nodes $v_i$ and $v_j$. Their major contributions lie in that they adopt the fusion technique of MRF and LDA to deal with community detection, which can well overcome the difficulties that two model’s parameters are hard to share and train together via factor graphs and belief propagation.

The emergence of factor graph models that integrate directed and undirected graphical models has greatly improved the performance of community detection. However, these probabilistic graph models generally adopt variational inference or Markov chain Monte Carlo (MCMC) sampling for model optimization, which inevitablely leads to high computational complexity. Deep learning, with the ability to effectively optimize on high-dimensional network data, has a potential in handling community detection.

%% file: deep_learning.tex
\section{Community Detection with Deep Learning}
\begin{table*}[htbp]
\centering
\caption{Summary of auto-encoder-based community detection, where "A" and “X” denote whether the approach utilize network topology and node attributes respectively, and ”-” represents no constraint.}
\label{aet}
\begin{tabular}{clccllll}
\toprule 
\bf Categories & \bf Approaches & \bf A & \bf X & \bf Encoder & \bf Decoder & \bf Focus & \bf Constraints\\
\midrule
\multirow{7}{*}{Stacked}      & Semi-DNR (2016) \cite{yang2016modularity}    & Yes  & No    & MLP    & MLP      
                              & Network embedding    & Pairwise constraint  
                              \\
                              \cline{2-8}
                               & DIR (2017) \cite{jin2017using}    & Yes  & Yes    & MLP    & MLP
                              & Network embedding    & -    \\
                                 \cline{2-8}
                              & INSNCCD (2018) \cite{cao2018incorporating}    & Yes  & Yes    & MLP    & MLP
                              & Network embedding    & Modularity maximization    \\
                              \cline{2-8}
                              & AAGR (2018) \cite{cao2018autoencoder}    & Yes  & Yes    & MLP    & MLP
                              & Network embedding    & Adaptive parameter    \\
                              \cline{2-8}
                              & CDDTA (2019) \cite{xie2019high-performance}    & Yes  & No    & MLP    & MLP
                              & Network embedding    & Regularization term    \\
                            \cline{2-8}           
                              & DeCom (2019) \cite{bhatia2019a}    & Yes  & No    & MLP    & MLP
                              & Clustering result    & Modularity maximization    \\
                               \cline{2-8}           
                              & NEC (2020) \cite{sun2020network}    & Yes  & Yes    & GCN    & Inner product
                              & Embedding and clustering    & Modularity maximization    \\ 
                              \hline
\multirow{4}{*}{Sparse}       & GraphEncoder (2014) \cite{tian2014learning}    & Yes  & No    & MLP    & MLP
                              & Network embedding    & Sparsity constraint    \\
                         \cline{2-8}        
                              & DFuzzy (2018) \cite{bhatia2018dfuzzy:}    & Yes  & No    & MLP    & MLP
                              & Clustering result    & \begin{tabular}[c]{@{}l@{}}Sparsity constraint and    \\ modularity maximization \end{tabular} 
                                                                                                      \\
                                                                                    \cline{2-8} 
                              & CDMEC (2020) \cite{xu2020stacked}    & Yes  & No    & MLP    & MLP
                              & Clustering result    & Sparsity constraint    \\ 
                              \hline
\multirow{2}{*}{Denoising}    & MGAE (2017) \cite{wang2017mgae:}    & Yes  & Yes    & GCN    & GCN
                              & Network embedding    & Interplay exploitation    \\
                            \cline{2-8}          
                              & GRACE (2017) \cite{yang2017graph}    & Yes  & Yes    & MLP    & MLP
                              & Embedding and clustering    & Propagation constraint    \\ 
                              \hline
\multirow{6}{*}{Variational}  & ARVGA (2018) \cite{pan2018adversarially}    & Yes  & Yes    & GCN    & Inner product
                              & Network embedding    & Prior constraint    \\
                              \cline{2-8} 
                              & VGAECD (2018) \cite{choong2018learning}    & Yes  & Yes    & GCN    & Inner product 
                              & Embedding and clustering    & -    \\
                              \cline{2-8}
                              & DAEGC (2019) \cite{wang2019attributed}    & Yes  & Yes    & GAT    & Inner product
                              &  Embedding and Clustering    & KL divergence constraint    \\
                            \cline{2-8}       
                               
                              & New VGAECD (2019) \cite{choong2019optimizing}    & Yes  & Yes    & GCN    & Inner product
                              & Embedding and clustering    & -    \\ 
                        \cline{2-8} 
                              & Ladder VAE (2019) \cite{DBLP:conf/aaai/SarkarMR20}  & Yes  & No   & GCN    & MLP
                              & Embedding and clustering    & -    \\                        \cline{2-8} 
                              & NetVAE (2019) \cite{jin2019network-specific}  & Yes  & Yes   & MLP    & MLP
                              & Network embedding    &  Prior constraint    \\ 
\bottomrule
\vspace{-0.4cm}
\end{tabular}
\end{table*}
In recent years, deep learning has drawn a great deal of attention and has been demonstrated to have great power on a wide variety of problems, including community detection. Classic deep learning explores and exploits convolutional neural networks (CNNs) and probability modeling
for community detection. For example, Sperl{\`{\i}} \emph{et al}. \cite{19} design a novel approach, based on CNNs and the topological characteristics of adjacency matrices, for automatic community detection. Sun \emph{et al}. \cite{9} propose a probabilistic generative model, i.e., vGraph, to jointly detect overlapping (and non-overlapping) communities and learn node (and community) representation. vGraph represents each node by a mixture of communities and defines a community as a Multinomial distribution over nodes.
Furthermore, Cavallari \emph{et al}. \cite{DBLP:conf/cikm/CavallariZCCC17} find that there is a closed-loop relationship among community detection, community embedding, and node embedding. Guided by this insight, they present a novel community embedding method, called ComE, to jointly solve the three tasks altogether. 
He \emph{et al}.\cite{DBLP:journals/tkde/HeLZWH21} design a novel self-translation network embedding (STNE) approach, which maps the content sequence to node identity sequence to improve community detection.

Although these methods have had reasonable performance on discovering communities, they are straightforward applications of deep learning to community detection \cite{20}, without considering the characteristics of networks, e.g., irregularity of network topology and complex network structures. In this section, we will discuss the following four types of methods that are designed for complex networks, i.e., auto-encoder-based, generative adversarial network-based, graph convolutional network-based, and methods integrating graph convolutional network and undirected graphical models.

\subsection{Auto-encoder-based Methods}
Auto-encoders \cite{hinton2006reducing} are simple but important neural models that convert high-dimensional (network) data into low-dimensional representations. Concretely, auto-encoders learn a new representation of data in an unsupervised manner using the encoder and decoder components. They always have multi-hidden layers and a symmetrical architecture, and the output of one layer is the input to its successive layer.
The objective of auto-encoders is to minimize the error between original input and reconstructed data to learn an optimal hidden representation, which can be denoted as:
\begin{equation}
\setlength\abovedisplayskip{8pt}
\setlength\belowdisplayskip{8pt}
\label{auto-encoder}
    Loss(\theta_1,\theta_2) = 
    \sum_{i=1}^{n}l(x_i,g(f(x_i;\theta_1);\theta_2)),
\end{equation}
where $f(\cdot;\theta_1)$  and $g(\cdot;\theta_2)$ are the encoder and decoder with parameters $\theta_1$ and $\theta_2$, and $l(\cdot)$ is the loss function. 

Herein, we choose several representative auto-encoder-based models for network community detection, and summarize their main characteristics 
in Table \ref{aet}. 
Since most auto-encoder-based methods derive network embeddings as their outputs (e.g.,  
\cite{yang2016modularity, cao2018autoencoder}),
clustering, such as K-means and spectral clustering, is subsequently applied to extract communities. An alternative is to integrate clustering  into the model (e.g., \cite{bhatia2018dfuzzy:,bhatia2019a}), 
to directly discover communities.
Depending on the type of auto-encoder used, we divide the models into four types, namely stacked, sparse, denoising and variational auto-encoders. 
Stacked auto-encoder, a basic type that consists of a series of auto-encoders, is often used as a block for other types of auto-encoders. Particularly, when a stacked model has other targets, such as sparsity and denoise, we classify them as sparse or denoising auto-encoders. 

\textbf{Stacked auto-encoders.} The semi-DNR \cite{yang2016modularity} stacks a sequence of auto-encoders to form a deep nonlinear reconstruction of the input networks (DNR), and requires that each layer of the encoder contains fewer neurons than the previous layer to reduce the data dimension and extract the most salient features in the input data. Semi-DNR makes full use of the prior knowledge if $v_i$ and $v_j$ belong to the same community to incorporate pairwise constraints between the two nodes in the network. 
Specifically, it defines a prior information matrix $O=(o_{ij})_{n\times n}$, where $o_{ij} = 1$ if $v_i$ and $v_j$ are known to be in the same community, or 0, otherwise. The loss function for semi-DNR is represented as:
\begin{equation}
\setlength\abovedisplayskip{8pt}
\setlength\belowdisplayskip{8pt}
    Loss = l(M,Z) + \lambda \operatorname{Tr}(H^{T}LH),
\end{equation}
where $L = D-O$, $\operatorname{Tr}(\cdot)$ is the trace of a matrix, $M$ the modularity matrix, $Z$ the reconstruction data, $H$ the representation matrix and $\lambda$ a parameter for making a trade-off between the reconstruction error 
and consistency of the new representation given the prior information. Further, 
a layer-wise stacked auto-encoder in DeCom \cite{bhatia2019a} is adopted to find seed nodes and add nodes to communities according to 
the structure of the network. It is remarkable that DeCom is suitable for handling large networks and there is no need to pre-define the number of communities due to the adaptive learning process. 
Besides, CDDTA \cite{xie2019high-performance} effectively combines transfer learning and auto-encoder. AAGR \cite{cao2018autoencoder} and DIR \cite{jin2017using} 
utilize stacked auto-encoders to incorporate the information of topology and attributes adaptively, thus
well realizing the balance between network topology and node attributes. NEC \cite{sun2020network} employs graph convolutional networks to encode and decode network data, which takes topology and attribute information as input, but only selects to reconstruct the adjacency matrix to ensure that the model can still work without node attributes.

\textbf{Sparse auto-encoders.} Large-scale networks are in general difficult to store and process, so it is necessary to have a sparse representation. A new line of research is to adaptively find the optimal representation by adding a sparse constraint to auto-encoder for this purpose.
GraphEncoder \cite{tian2014learning} introduces an explicit regularization term for the hidden layer to restrict the size of hidden representation. If $z_i$ is the $i$-th vector of reconstructed data,
the reconstruction error with sparsity constraints are as follows:
\begin{equation}
\setlength\abovedisplayskip{8pt}
\setlength\belowdisplayskip{8pt}
    Loss = \sum_{i=1}^{n}\left \| z_i - x_i\right \|_2 + \beta KL(\rho|\widehat{\rho}),
\end{equation}
where $\beta$ controls the sparsity penalty, $\rho$ and $\widehat{\rho}$ are the sparsity parameters, where the former denotes the average activation of a neuron across a collection of training samples, and the latter denotes the average activation across all training samples. 
The time complexity of GraphEncoder is $O(nbd)$, where $b$ is the maximum number of hidden layer nodes, and $d$ is the average degree of the graph.
DFuzzy \cite{bhatia2018dfuzzy:} is a parallel and scalable fuzzy clustering model with sparse auto-encoders as building blocks. 
It trains an auto-encoder using personalized PageRank, which is effective for capturing relationships among network nodes. Besides, CDMEC \cite{xu2020stacked} combines transfer learning with auto-encoder, where input matrix $A$ is used to build four similarity matrices of complex networks. CDMEC takes one matrix as the source domain, and the other three matrices as the target domain to obtain multiple distinct low-dimensional feature representations. 
All representations are then put into a clustering algorithm,
and the clustering results are integrated
into a new, concensus matrix $Q$. The consensus matrix $Q$ is introduced to measure the co-occurrence of samples in the clustering result, where $Q_{ij}$ represents the average times that $v_i$ and $v_j$ are grouped into the same class. 

\textbf{Denoising auto-encoders.} Denoising auto-encoders can be applied to noisy inputs to get node representation that is robust to noise. 
MAGE \cite{wang2017mgae:} first employs a convolutional network to integrate content and structure information, and then iteratively adds random noises to content information in the auto-encoder process. In this way, the structure information and content information are integrated into a unified framework, and the interplay between the two can
be analyzed. Further, Yang \emph{et al}. \cite{yang2017graph} propose GRACE
to deal with dynamic networks. They model clusters under the consideration of network dynamics, and believe that the formation of clusters requires dynamic embedding to reach a stable state.

\textbf{Variational auto-encoders.} There are also approaches based on variational auto-encoder \cite{DBLP:journals/corr/KingmaW13}, which views the hidden representation as a latent variable with its own prior distribution. In variational inference, it exploits an approximation $q(H|X)$ of the true posterior $p(H|X)$ of the latent variable and tries to approximate the variational posterior $q(H|X)$ to the true prior $p(H)$ using the KL-divergence as a measure.
For instance, the theme of ARVGA \cite{pan2018adversarially} is not only to minimize the reconstruction errors of network structure, but also to enforce the latent codes to match a prior distribution:
\begin{equation}
\setlength\abovedisplayskip{8pt}
\setlength\belowdisplayskip{8pt}
    Loss = \mathbb{E}_{q(H|(X, A))}[\log p(\widehat {A}|H)] - KL[q(H|X,A) || p(H)].
\end{equation}
During the training of VGAECD \cite{choong2018learning}, the reconstruction loss deviates from its primary objective of clustering.
The new VGAECD \cite{choong2019optimizing} rectifies this issue by introducing  a dual variational objective. Further,
Sarkar \emph{et al}. \cite{DBLP:conf/aaai/SarkarMR20} propose a gamma ladder VAE based deep generative model that infers multilayered embeddings for the nodes via multiple layers of stochastic latent variables to improve the performance of community detection.
Jin \emph{et al}. \cite{jin2019network-specific} propose the NetVAE that uses one encoder and one dual decoder with two different generative mechanisms to reconstruct network topology and node attributes separately. 

\subsection{Generative Adversarial Network-based Methods}
Generative adversarial networks (GANs) \cite{5}, which are inspired by the minimax two-player game, have achieved unprecedented success in various fields.
GANs typically consist of two modules, a
generator $\mathcal{G}$ and a discriminator $\mathcal{D}$. The generator is to capture the data distribution, i.e., to generate samples that are as similar to the real data as possible; while the discriminator is to estimate the probability that a sample a piece of real data rather than synthetic data generated by the generator. 
Formally, the training process of GANs can be defined as:
\begin{align}
\setlength\abovedisplayskip{8pt}
\setlength\belowdisplayskip{8pt}
    \min \limits_{\mathcal{G}} \max \limits_{\mathcal{D}} \mathcal{V} (\mathcal{G}, \mathcal{D}) = & \min \limits_{\mathcal{G}} \max \limits_{\mathcal{D}} (\mathbb{E}_{x\sim p_{data(x)}} [\log \mathcal{D}(x)] 
    \notag
    \\ &+ \mathbb{E}_{z\sim p_{z(z)}} [\log (1 - \mathcal{D}(\mathcal{G}(z)))] ),
\end{align}
where the first expectation is the loss of discriminator for real data and the second is the loss of discriminator for synthetic data generated by the generator.

The inspiration of applying GANs to community detection came from the fact that GANs are usually unsupervised, and (in theory) the new data generated have the same distribution as real data, which provides a powerful network data analysis capability.
Jia \emph{et al}. \cite{1} propose a novel method called CommunityGAN to adopt the idea of affiliation graph model (AGM) to boost the performance by introducing the minimax competition between the network motif-level generator and the discriminator. It first composes some representation vectors of nodes by assigning each node-community pair a nonnegative factor that represents the degree of membership of the node to community, and then optimizes such representation through a specifically designed GAN to detect communities. The joint value function is formulated as: 
\begin{align}
    \min \limits_{\theta_\mathcal{G}} \max \limits_{\theta_\mathcal{D}} & \mathcal{V} (\mathcal{G},\mathcal{D}) = \sum_{i=1}^n (\mathbb{E}_{m\sim p_{true(.|v_i)}} [\log \mathcal{D}(m; \theta_\mathcal{D})] 
    \notag
    \\ &+ \mathbb{E}_{s\sim \mathcal{G}_{(s|v_i;\theta_\mathcal{G})}} [\log (1 - \mathcal{D}(\mathcal{G}(s; \theta_\mathcal{D})))] ),
\end{align}
where $\theta_\mathcal{D}$ (and $\theta_\mathcal{G}$) is the union of
representation vectors of all nodes in the discriminator $\mathcal{D}$ (and generator $\mathcal{G}$), $m$ the motifs of networks and $s$ the subset of nodes. By employing GANs, CommunityGAN can find overlapping communities and learn a graph representation altogether. Further, 
Zhang \emph{et al}. \cite{DBLP:conf/kdd/ZhangXYLWZY20} present a novel approach of seed expansion with generative adversarial learning (SEAL). It employs a discriminator to predict whether a community is real or not and a generator to construct communities to trick the discriminator by implicitly fitting features of real ones for learning heuristics for community detection.
Tao \emph{et al}. \cite{DBLP:conf/ijcai/TaoLLW019} propose an adversarial graph auto-encoder (AGAE) method, which incorporates ensemble clustering into a deep graph embedding process to guide the network training utilizing an adversarial regularizer.

There are also methods based on GANs to derive node representation that can be applied to community detection, e.g., employing clustering algorithms such as K-means for deriving embeddings to acquire resulting communities \cite{DBLP:conf/ijcai/SunWHTH19,DBLP:conf/kdd/GaoPH19a,DBLP:journals/tnn/Hong0W20}. 
He \emph{et al}. \cite{3} further argue that the existing GANs-based methods do not make full use of the essential advantages of GANs, which are to  learn the underlying representation mechanism rather than the representation itself. To this end, they propose to utilize adversarial idea on the representation mechanism
to acquire node representation for downstream tasks.
Specifically, 
the training loss is defined as follows:
\begin{align}
\min \limits_{\mathcal{E,G}} \max \limits_{\mathcal{D}} & \mathcal{V} (\mathcal{G}, \mathcal{D}, \mathcal{E})
   = \mathbb{E}_{x\sim p_{data(x)}} [\log \mathcal{D}(MI(x,\mathcal{E}(x)))] 
    \notag
    \\ &+ \mathbb{E}_{z\sim p_{z(z)}} [\log (1 - \mathcal{D}(MI (\mathcal{G}(z),z)))] ),
\end{align}
where $\mathcal{E}$ represents the encoder that derives node representation, $MI(x,\mathcal{E}(x))$ is the mutual information between the node attributes and node representation, $\mathcal{D}$ is the discriminator that identifies the mutual information from either positive or negative samples, and $\mathcal{G}$ is the generator that generates negative samples by calculating the mutual information between fake node attributes based on Guassian noise. Yang \emph{et al}. \cite{4} argue that most GANs compare the results of embeddings with samples obtained from Gaussian distribution without rectification from real data, making them not truly beneficial for adversarial learning. Therefore, they design a joint adversarial network embedding (JANE) model, which jointly distinguishes the real and fake combinations of embeddings, topology information and node attributes, to improve node embeddings and the performance of network analysis. 

\subsection{Graph Convolutional Network-based Methods}
Graph convolutional networks (GCNs) \cite{DBLP:conf/iclr/KipfW17}, the most representative branch of \textit{graph neural network} methods\cite{wu2020comprehensive} for learning representation from graph data, 
have attracted a great deal of attention thanks to its success on supervised and semi-supervised classification of nodes in a network. 
Several novel GCNs-based algorithms have also been developed lately to exploit the power of GCNs for effectively modeling and inferring high-dimensional complex network data for community detection.

Jin \emph{et al}. \cite{17} raise the concern that embeddings derived from GCNs are not community-oriented and community detection is inherently unsupervised. To address this problem, they introduce an unsupervised model, named JGE-CD, for community detection through joint GCN embedding. It consists of three modules, a dual encoder that derives two embeddings using the original attribute network and its variant; a community detection module that stacks on top of dual encoder to detect community; and a topology reconstruction module that is employed to reconstruct network topology. Formally, the probability that the $i$-th node belongs to the $r$-th community is defined as:
\begin{equation}
u_{ir}= \frac{\operatorname{exp}(\theta_r^Th_i)}{\sum_{r=1}^{k}\operatorname{exp}(\theta_{r}^Th_i)},
\end{equation}
where $h_i$ represents the embedding of node $v_i$ obtained from GCN and $\theta$  the model parameters.
Furthermore,
He \emph{et al}. \cite{13} extend JGE-CD by designing a new GCN approach that casts MRFasGCN (to be discussed in Section 4.4 shortly) as an encoder, and exploits a community-centric dual encoder to reconstruct network topology and node attributes separately, so as to perform unsupervised community detection. In particular, the decoder for reconstructing network topology is denoted as:
\begin{equation}\label{topology}
\widehat{A}=\operatorname{sigmoid}\left(DU W U^{T} D^{T}\right),
\end{equation}
where $U$ is the probability distribution matrix of nodes belonging to different communities derived from the encoder, $D$  the node degree matrix and $W$  the weight matrix of neural networks. The decoder for reconstructing attributes is inspired by topic modeling, i.e., nodes in the same community are more likely to have similar distributions of attribute words. The attribute matrix can be generated by:
\begin{equation}
\widehat{X}=U \cdot R,
\end{equation}
where the definition of $U$ is the same as that in \eqref{topology} and $R$ is the probability matrix of communities selecting attribute words from the entire word set.

More recently, some researches for community detection make use of GCNs on heterogeneous networks that contain a diversity of types of nodes and relationships. Zheng \emph{et al}. \cite{16} design a heterogeneous-temporal GCN, namely HTGCN, to detect community from hetergeneous and temporal networks. Concretely, it first obtains feature representation of each hetergeneous network at each time step by adopting a heterogeneous GCN, and then utilizes a residual compressed aggregation mechanism to express both the static and dynamic characteristics of community. 
Beyond that, there are also certain approaches incorporating graph convolutional network with  undirected graphical models, e.g., MRFasGCN \cite{6} and GMNN \cite{7}, which will be discussed next.

\subsection{Integrating Graph Convolutional Network and Undirected Graphical Models}
In the last few years, a number of studies have begun to integrate graph convolutional network (GCN) and undirected graphical models (e.g., MRF or CRF) for community detection. 
The main idea of this line of research is that GCN essentially constructs node embeddings through local feature smoothing, which does not consider community properties and makes the node embeddings not community-oriented. While undirected graphical models generally offer a good global objective to describe community, it does not consider information on nodes and requires a substantial amount of computation for learning the model. Therefore, GCN and undirected graphical models are complementary and can be combined to take advantage of their strengths.

A major work in this line is MRFasGCN \cite{6}, 
which integrates GCN with MRF to solve the problem of semi-supervised community detection in attributed networks. The method first extends NetMRF (as discussed in Section 3.2) to extended MRF (eMRF) by adding both unary potentials and attribute information, and then reparameterizes the MRF model to make it fit to the GCN architecture. 
The energy function of eMRF is defined as:
\begin{equation}
E(C;A,X) = \sum_{v_i} - p(h_i^{c_i}) + \alpha \sum_{v_i\neq v_j}\mu (c_i,c_j)\eta (v_i,v_j),
\end{equation}
where 
$- p(h_i^{c_i})$, whose value comes from the result of GCN, denotes the unary potential representing the probability that node $v_i$ belongs to community $c_i$,
$\mu (c_i,c_j)\eta(v_i,v_j)$ is the pairwise potential where $\mu (c_i,c_j)$ represents the similarity relationship between communities of nodes $v_i$ and $v_j$, and $\eta(v_i,v_j)$ is the similarity of attributes
of nodes $v_i$ and $v_j$, and $\alpha$ is the parameter for making a tradeoff between the unary and pairwise potentials.
The time complexity of MRFasGCN is $O(mqlk{^2})$, where $q$ is the dimension of node attribute and $l$ is the number of hidden units of the first layer.


After MRFasGCN, several other lines of work incorporate MRF or CRF into GCN to learn node embedding for community detection. 
Qu \emph{et al}. \cite{7} propose a new approach, called graph Markov neural network (GMNN),  combines the advantages of both statistical relational learning and graph neural networks. A GMNN is able to learn an effective node representation and model label dependency between different nodes, thereby completing the task of semi-supervised node classification. The model parameters can be learned by employing pseudolikelihood variational expectation-maximization 
\cite{neal1998view} to optimize the evidence lower bound 
(ELBO) 
of log-likelihood function, which is formulated as:
\begin{equation}
\begin{split}
&\log p_{\phi}(y_{L}| x_{V}) \geqslant \\  
&E_{q_{\theta}}(y_{U}| x_{V})[\log p_{\phi}(y_{L},y_{U}|x_{V}) - \log q_{\theta}(y_{U} | x_{V})],      
\end{split}
\end{equation}
where $\log p_{\phi}(y_{L}|x_{V})$ is the log-likelihood function of observed node labels, and $q_{\theta}(y_{U} | x_{V})$ is any distribution over $y_U$. Noting that the equality holds when $q_{\theta}(y_U | x_V) = p_{\phi}(y_{U} | y_{L}, x_{V})$. Gao \emph{et al}. \cite{8} find that the existing GCNs fail to preserve the similarity relationship between different nodes hidden in network data. To handle this issue, they add a CRF layer to GCNs to force similar nodes to have similar hidden features. This will enhance the quality of node embeddings and, in turn, improve the performance of network analysis.

\subsection{Summary for deep learning}
Deep learning-based methods usually realize community detection by designing a community-oriented or universal network representation, followed by some clustering algorithms. Different types of deep-learning methods have been adopted to derive network representations, e.g., GCN or GAN. Learning such representations is to map network data from a high-dimensional input space to a low-dimensional feature space, with the superiority of low computational complexity and effective data fusion. However, similar to the most probabilistic graphical models, the existing deep learning-based methods are more suitable for community structures with homophily (where the nodes in a community are densely connected whereas nodes in different communities are sparsely linked), which may limit the robustness of these models.



%% file: Application.tex
\section{Applications of Community Detection}
We start our discussion with a summary of the benchmark datasets that have been used in the area of community detection. We then describe real applications of community detection in many application fields.
\subsection{Open Datasets}
We have put the detailed information of the datasets used for community detection in a publicly accessible web\footnote{http://bdilab.tju.edu.cn/} to facilitate open research on this rapidly-developing topic. These datasets can be  separated 
into two groups, synthetic networks and real-world networks. 


\subsubsection{Synthetic networks}
There are two classes of randomly generated synthetic networks with known community structures, i.e., the Girvan-Newman (GN) \cite{2002Girvan} and LFR networks \cite{Lancichinetti}. The GN network consists of four non-overlapping communities with the same size. Each community has 32 nodes, each of which connects with 16 other nodes on average. Among these 16 edges, $Z_{in}$ edges connect to nodes of the same community, $Z_{out}$ edges to nodes of different communities, and $Z_{in}+ Z_{out}=16$.
The LFR network, another widely adopted benchmark
for testing the performance of algorithms for community detection, has distributions of node degree and community size which follow power laws with tunable exponents. The LFR network captures several important features of real-world systems, e.g., the scale free property.

\subsubsection{Real-world Networks}
The real-world networks that we examine include four types, i.e., social networks, citation networks, collaboration networks, and others, listed in Table 5. To be specific, social networks are formed by individuals and their interactions, including eight representative datasets such as Football and DBLP (Table 5). Citation networks consist of papers (or patents) and their relationships (e.g., citation or inclusion), including eight classic datasets such as Cora and arXiv. Collaboration networks are comprised of scientists and their collaborations (i.e., co-authoring papers), including four typical datasets such as Computer Science and Medicine. The nodes in these networks range from tens to millions, and the maximum number of edges reaches hundreds to hundreds of millions.

\begin{table}[htbp]
    \setlength{\abovecaptionskip}{1.5pt}
    \setlength{\belowcaptionskip}{0pt}
    \centering
    \caption{The statistics of real-world networks.}
    \scalebox{0.8}{
    \begin{tabular}{ m{1.9cm}<{\centering} m{3.0cm} m{1.7cm} m{1.7cm}}
    \toprule
    \bf Categories & \bf Datasets 
    & \bf \#Nodes & \bf \#Edges  \\
    \midrule
	
    \multirow{7}*{\shortstack{Social\\Networks}} 
    {} & Friendship7 \cite{2013Overlapping} & 68 & 220  \\
    \cline{2-4}
    {} & Football \cite{2002Girvan} & 115 & 613  \\
    \cline{2-4}
    {} & Facebook \cite{DBLP:conf/nips/McAuleyL12} & 1,045 & 26,749 \\
    \cline{2-4}
    {} & LiveJournal \cite{2014Community} & 44,093 & 871,409  \\
     \cline{2-4}
    {} & Twitter \cite{DBLP:conf/kdd/ZhangXYLWZY20} & 87,760 & 1,293,985 \\
     \cline{2-4}
    {} & Orkut \cite{2014Community} & 297,691 & 7,747,026 \\
     \cline{2-4}
    {} & DBLP \cite{DBLP:journals/tkde/LierdeCC20} & 317,080 & 1,049,866 \\
    \cline{2-4}
    {} & Youtube\cite{DBLP:journals/tkde/LierdeCC20} & 1,134,890 & 2,987,624 \\
    \hline
      
    \multirow{8}*{\shortstack{Citation\\Networks}} 
	{} & Small-hep \footnotemark[2] & 397 & 812   \\
    \cline{2-4}
	{} & Polblogs \cite{DBLP:conf/kdd/AdamicG05} & 1490 & 16718   \\
     \cline{2-4}
    {} & Cora \cite{2009Simple} & 2708 & 5429  \\
	 \cline{2-4}
	{} & Citeseer  \cite{2009Simple} & 3312 & 4732  \\
    \cline{2-4}
	{} & Large-hep \footnotemark[2] 
     & 11,752 & 134,956  \\
	 \cline{2-4}
	{} & Pubmed \cite{DBLP:journals/aim/SenNBGGE08} & 19,729 & 44,338  \\
	 \cline{2-4}
	{} & arXiv \cite{Ginsparg2011ArXiv} & 576,000 & 6,640,000  \\
	 \cline{2-4}
	{} & US patents \cite{DBLP:conf/kdd/LeskovecKF05} & 3,700,000 & 16,500,000  \\
	\hline
	 \multirow{4}*{\shortstack{Collaboration \\Networks}}
	 {} & Computer Science \cite{12}& 22,000 & 96,800  \\
	 \cline{2-4}
	 {} & Chemistry \cite{12} & 35,400 & 157,400  \\
	 \cline{2-4}
	 {} & Medicine \cite{12} & 63,300 & 810,300  \\
	 \cline{2-4}
	 {} & Engineering \cite{12} & 14,900 & 49,300  \\
	 \hline
	\multirow{2}*{\shortstack{Others}} 
	{} & Amazon \cite{DBLP:journals/tkde/LierdeCC20} & 334,863 & 925,872  \\
	 \cline{2-4}
	{} & Google \cite{DBLP:journals/im/LeskovecLDM09} & 875,000 & 4,320,000 \\
    \bottomrule
    \end{tabular}
    }
    \label{tab:dataset}
    \vspace{-0.5cm}
\end{table}
\footnotetext[2]{https://www.cs.cornell.edu/projects/kddcup/datasets.html}

\subsection{Practical Applications}
We first discuss the applications of community detection on different domains, and then extend to other network analysis tasks. We finish by discussing the potential of community detection to network science.

\subsubsection{Applications in Different Areas}
Community detection has diverse applications across different domains \cite{DBLP:journals/pvldb/ChuWPZYC19}, such as online social networks, neuroscience and image understanding. 
Online social networks, including Facebook, Twitter and Wechat, comprise the interactions among people through the web. Discovering community in such networks is an effective way to infer the relationships of individuals, which has been adopted for tasks such as spammer detection and crisis response. 
Yin \emph{et al}. \cite{DBLP:conf/icde/YinHZWZHS16} develop a unified probabilistic generative model, namely user-community-geo-topic (UCGT), based on a Bayesian network that infers users' social communities by incorporating spatio-temporal data and semantic information to improve the accuracy and interpretability of social community detection.
Wu \emph{et al}. \cite{10} design a novel MRFwithGCN model,
which introduce a MRF layer that captures user following information to refine prediction made by GCN for social spammer detection.  

Neuroscience is a discipline studying the nervous systems and brain. With the recent development of brain mapping and neuroimaging techniques,
the brain has begun to be modeled as networks. A large amount of effort has been put forward to exploit such networks to help extract the functional subdivisions of the brain. Liu \emph{et al}. \cite{15} propose a framework of siamese community preserving graph convolutional network (SCP-GCN). The method first retains the community structure considering the intra-community and inter-community properties in learning process, and then uses siamese architecture that models the pair-wise similarity to guide this learning process.
Jin \emph{et al}. \cite{2019Multiscale} argue that the existing studies typically construct community structures of brain networks employing resting-state functional magnetic resonance imaging (fMRI) data.
They introduce the dynamic time warping (DTW) algorithm that analyzes the synchronization and asynchronism of fMRI time series to extract the correlation between brain regions. 

Image understanding is to generate semantic descriptions of images. Recently, works on image understanding utilizing community have emerged. 
Li  \emph{et al}. \cite{DBLP:journals/pami/LiTM19} present a novel deep collaborative embedding (DCE) model, which learns knowledge from weakly supervised community-contributed resources, for multiple image understanding tasks simultaneously. Li \emph{et al}. \cite{DBLP:journals/tnn/LiTH18} propose a novel semi-supervised RSNMF model by explicitly exploring the block-diagonal structure of data for image representation.


\subsubsection{Promotion of Network Analysis}
With the great success of community detection, numerous application problems, e.g., recommendation and link prediction, have been formulated as finding community structures in network systems. We now discuss how the existing community detection methods are utilized to solve some of these problems.

Recommendation is a common task that addresses the issue of information overload for users by establishing a profile of user interests based on items in their purchasing or browsing history and later recommending similar items to users. The existing methods for recommendation including
collaborative filtering \cite{DBLP:conf/ijcai/0007YTZT19} and neural networks \cite{DBLP:conf/sigir/0001DWLZ020}. In particular, the concept of community has been employed to improve the quality of recommendation. Eissa \emph{et al}. \cite{DBLP:conf/www/EissaEM18} make recommendation based on interest-based communities that generated from topic based attributed social networks. 
Satuluri \emph{et al}. \cite{DBLP:conf/kdd/SatuluriWZQWDTJ20} present a general-purpose representation layer, i.e., similarity-based clusters (SimClusters), which settles a multitude of recommendation tasks at Twitter via detecting bipartite communities from the user-user network and leveraging them as a representation space.

Link prediction is another important
task in network mining. It deals with missing connections and predicts possible connections in the future through the analysis of observed network structure and external information. A large number of approaches have been proposed to facilitate link prediction via considering community. Xu \emph{et al}. \cite{DBLP:conf/www/XuWCY17a} indicate that the existing metric learning for link prediction ignores community that contain abundant structural information. Therefore, they design community specific similarity metrics by means of joint community detection to deal with cold-start
link prediction where edges between nodes are unavailable. De \emph{et al}. \cite{DBLP:journals/tkde/DeBSGC16} propose a stacked two-level learning framework, which first learns a local similarity model exploiting locality structures and node attributes, and then combines the model with community-level features that derive utilizing co-clustering for link prediction.

\subsubsection{Promotion of Network Science}
Network science is an interdisciplinary research area, which can be used not only in computer science, but also in other fields such as sociology and biology. Community detection, one of the most important problem in network analysis, can tremendously promote the development of network science. For example, in citation network, nodes represent papers and edges represent the citations among papers. By grouping the papers (i.e., discovering communities where papers have similar attributes such as belonging to the same author or topic), we can analyze the influence of authors and accurately grasp the latest research trends or technologies, which is of guiding significance for comprehending network and further analyzing network pattern \cite{DBLP:journals/dint/WanZZT19}. Similarly, in sociology and biology, community detection also provides a deeper understanding of network structure and promotes its development both in academia and industry.

%% file: challenges.tex

%% file: future.tex
\section{Future Directions}
While learning-based community detection, including probabilistic graphical model and deep learning, has demonstrated superior performance across a variety of problems and domains, there are challenges that need to be addressed. In this section, we briefly discuss these challenges and future research directions potentially worth pursuing.
\subsection{Large Networks}
With the rapidly increasing scale of
network data, more large networks have become the standard across many different scientific domains. These networks typically have tens of thousand or billions of nodes and edges as well as complex structural patterns. Most existing community detection methods face excessive constraints on such large networks due to the potentially prohibitive demand on memory and computation. They may require a large number of training instances \cite{DBLP:conf/aaai/JinWDHZ16} or model parameters \cite{DBLP:conf/aaai/WangJMD19} to make the existing methods effective. Moreover, the existing approaches typically
handle these problems by network reduction 
or  approximation, 
which may lose some important network information and affect the modeling accuracy. This raises the question of how to devise a framework that far exceeds the current benchmarks in accuracy and efficiency.

\subsection{Community Interpretability}
Although community detection has been studied for more than a decade, the interpretability of community remains an important and critical issue to be adequately addressed.
Most current community detection methods utilize top ranked words or short phrases in the results to summarize communities, even though the attribute information of nodes is typically complete sentences that have more information than individual words \cite{DBLP:journals/pvldb/CaiZZCH17,DBLP:conf/aaai/HeFJWZ17}.
However, these methods may not be intuitive enough for understanding the semantics of communities due to the small number of words and unclear relationship between words. How to make the best use of network information to provide a better semantic interpretation for community is one of future research directions.

\subsection{Adaptive Community Model Selection}
Adaptive model selection for community detection aims to choose the most appropriate algorithm for discovering community, according to the characteristics of different networks (e.g., heterogeneous or dynamic) or specific requirements of different tasks (e.g., the highest accuracy or the lowest time complexity). Although the existing methods can be extended from one network or task to another to some extent (which inevitably affects the accuracy and stability of the resulting model) \cite{DBLP:conf/ijcai/ShaoZYWZY19}, \cite{DBLP:conf/aaai/LiSHZ18}, 
few of them consider how to perform model adaptation. Thus, focus has shifted to designing a unified architecture that can automatically adapt to specific tasks or networks while maintaining model accuracy and stability instead of proposing diverse frameworks for different networks or tasks. This is an emerging research area that would be challenging but rewarding.

\subsection{Networks with Complex Structures}
Many real-world networks are heterogeneous, dynamic, hierarchical, or incomplete. Heterogeneous networks \cite{DBLP:conf/www/LiWEKWZ17} are those that contain different types of nodes and edges, or different types of descriptions on nodes and edges, such as text and images. Dynamic networks \cite{DBLP:conf/icdm/DiTursiGB17} are networks whose topology and/or attributes change over time. Dynamic networks appear when nodes and edges are added or deleted, thus altering the properties of nodes or edges. Hierarchical networks \cite{DBLP:conf/kdd/ChenTXYH16} are composed of several layers, each of which has specific semantics and functions. Incomplete networks \cite{DBLP:conf/www/LinKYWJL12} are the ones with missing information of their topology,  nodes, or edges.
While these networks can be partly explored by the learning-based community detection, there still exist several serious issues. First, most existing methods assume homogeneous networks, which may in fact difficult to handle. Second, due to the variability of dynamic networks, most existing methods, especially the ones based on deep learning, need to be re-trained over a series of steps when the networks evolve, which is very time consuming and may not meet the real time processing demand. Third, hierarchical networks typically have different types of relationships across the network hierarchies, which are important while often not well handled by the existing methods. Moreover, almost all the existing methods regard the networks  to be analyzed to be complete and accurately documented without noise. Unfortunately, this is rarely the case in practice as it is challenging to obtain complete information of the networks. Therefore, new methods should be developed to handle these issues to better improve the performance of community detection on these types of complex networks.

\subsection{Integrating Statistical Modeling and Deep Learning}
Although several methods have been proposed to combine statistical modeling with deep learning, such as MRFasGCN, it is still a virgin but promising research area. For instance, the existing methods typically utilize the prior knowledge (e.g., communities) that statistical model offers to refine the embeddings of GCN to improve resulting communities. However, these methods may not fully consider the time complexity or interpretability of the models, raising enormous challenges to community detection in practice. 
In addition, the community patterns in the real world networks are usually diverse, e.g., community structure with heterophily (where the nodes in different communities are closely connected while the nodes in the same community are sparsely connected) or 
randomness (where the edges between nodes are more likely to be randomly generated), it is essential to design new robust methods by integrating statistical modeling and deep learning \cite{6}, thereby detecting the community structure in the network more accurately. Furthermore, it remains an open problem to integrate statistical modeling in deep learning methods. For example, it is difficult to apply a strategy for recommendation or medical diagnosis to make deep learning become better representation learning, which in turn facilitates more accurate recommendation or diagnose. New innovative algorithms are highly desirable to integrate statistical inference and deep learning to help deep learning more produce interpretable network representation models that are  suitable for various network problems in broad application fields.